
\input harvmac
\def\ibar{{\bar\imath}}
\def\jbar{{\bar\jmath}}
\noblackbox

\lref\CHSW{P. Candelas, G. Horowitz, A. Strominger and E. Witten,
``Vacuum Configurations for Superstrings,'' Nucl. Phys. {\bf B258}, 46
(1985).}

\lref\Sen{A. Sen, ``A Note on Enhanced Gauge Symmetries in M and
String Theory,'' JHEP {\bf 9709}, 001 (1997) [arXiv:hep-th/9707123].}

\lref\Andy{A. Strominger, ``Superstrings with Torsion,''
Nucl. Phys. {\bf B274}, 253 (1986).}

\lref\freycos{A. Frey and A. Mazumdar, ``Three-form Induced Potentials,
Dilaton Stabilization, and Running Moduli,'' [arXiv:hep-th/0210254]\semi
S. Kachru, R. Kallosh and S. Trivedi, unpublished.
}

\lref\buscher{T.H. Buscher, ``A Symmetry of the String Background
Field Equations,'' Phys. Lett. {\bf B194}, 59 (1987)\semi
T.H. Buscher, ``Path Integral Derivation of Quantum Duality in
Nonlinear Sigma Models,'' Phys. Lett. {\bf B201}, 466 (1988).}

\lref\TayVaf{T. Taylor and C. Vafa, ``RR Flux on Calabi-Yau
and Partial Supersymmetry Breaking,'' Phys. Lett. {\bf B474},
130 (2000) [arXiv:hep-th/9912152].}

\lref\GidOl{S. Giddings and O. DeWolfe, ``Scales and Hierarchies
in Warped Compactifications and Brane Worlds,''
[arXiv:hep-th/0208123].}

\lref\Greene{B. Greene, K. Schalm and G. Shiu, ``Warped Compactifications
in M and F Theory,'' Nucl. Phys. {\bf B584}, 480 (2000)
[arXiv:hep-th/0004103].}

\lref\SYZ{A. Strominger, S.T. Yau and E. Zaslow, ``Mirror Symmetry
is T-Duality,'' Nucl. Phys. {\bf B479}, 243 (1996)
[arXiv:hep-th/9606040].}

\lref\DasBeck{K. Becker and K. Dasgupta, ``Heterotic Strings with
Torsion,'' [arXiv:hep-th/0209077]\semi
K. Becker, M. Becker, K. Dasgupta and P. Green, to appear.
}

\lref\Gukovun{S. Gukov, private discussions.}

\lref\Atish{
A.~Dabholkar and C.~Hull,
``Duality Twists, Orbifolds, and Fluxes,''
[arXiv:hep-th/0210209].
}

\lref\Janetal{S. Gurrieri, J. Louis, A. Micu and D. Waldram, ``Mirror
Symmetry in Generalized Calabi-Yau Compactifications,'' 
[arXiv:hep-th/0211102].}

\lref\Humbgrp{G. Cardoso, G. Curio, G. Dall'Agata, D. L\"ust, P. Manousselis
and G. Zoupanos, ``Non-K\"ahler String Backgrounds and their Five Torsion
Classes,'' [arXiv:hep-th/0211118].}

\lref\phenom{J. Lykken, E. Poppitz and S. P. Trivedi,
``Branes with GUTS and supersymmetry breaking," Nucl. Phys. {\bf B 543}, 105 (1999)
[arXiv:hep-th/9806080] \semi G. Aldazabal, S. Franco, L. Ibanez, R. Rabadan and
A. Uranga, ``$D=4$ Chiral String Compactifications from Intersecting
Branes,'' J. Math. Phys. {\bf 42}, 3103 (2001) [arXiv:hep-th/0011073]\semi
D. Berenstein, V. Jejjala and R. Leigh, ``The Standard Model on a
D-brane,'' Phys. Rev. Lett. {\bf 88}, 071602 (2002) [arXiv:hep-th/0105042]\semi
M. Cvetic, G. Shiu and A. Uranga, ``Three Family Supersymmetric Standard-like
Models from Intersecting Brane Worlds,'' Phys. Rev. Lett. {\bf 87}, 201801
(2001) [arXiv:hep-th/0107143]\semi
D. Cremades, L. Ibanez and F. Marchesano, ``Standard Model
at Intersecting D5-Branes: Lowering the String Scale,''
Nucl. Phys. {\bf B643}, 93 (2002) [arXiv:hep-th/0205074]\semi
R. Blumenhagen, V. Braun, B. Kors and D. L\"ust, ``Orientifolds
of K3 and Calabi-Yau Manifolds with Intersecting D-Branes,''
JHEP {\bf 0207}, 026 (2002) [arXiv:hep-th/0206038]\semi
A. Uranga, ``Local Models for Intersecting Brane Worlds,''
[arXiv:hep-th/0208014]\semi
C. Kokorelis, ``Exact Standard Model Structures from Intersecting
Branes,'' [arXiv:hep-th/0210004]\semi
D. Bailin, G.V. Kraniotis and A. Love, ``Standard-like Models
from Intersecting D5 Branes,'' [arXiv:hep-th/0210219]\semi
R. Blumenhagen, L. Goerlich and T. Ott, ``Supersymmetric
Intersecting Branes on the Type IIA $T^6/Z_4$ Orientifold,''
[arXiv:hep-th/0211059].
} 

\lref\micu{J. Louis and A. Micu, ``Type II Theories Compactified on 
Calabi-Yau Threefolds in the Presence of Background Fluxes,'' 
Nucl. Phys. {\bf B635}, 395 (2002) [arXiv:hep-th/0202168].}

\lref\ferretti{R. Argurio, V. Campos, G. Ferretti and R. Heise,
``Freezing of Moduli with Fluxes in Three-Dimensions,'' Nucl. Phys.
{\bf B640}, 351 (2002) [arXiv:hep-th/0205295].}

\lref\Essays{{\it Mirror Symmetry I}, S.T. Yau, ed., American
Mathematical Society, Providence, USA (1998).}

\lref\witsup{E. Witten, ``Nonperturbative Superpotentials in String
Theory,'' Nucl. Phys. {\bf B474}, 343 (1996) [arXiv:hep-th/9604030].}

\lref\DFhananykol{A.~Hanany and B.~Kol, ``On orientifolds, discrete
torsion, branes and M theory,'' JHEP {\bf 0006}, 013 (2000)
[arXiv:hep-th/0003025].
}

\lref\kst{S. Kachru, M. Schulz and S. P. Trivedi, ``Moduli Stabilization
from Fluxes in a Simple IIB Orientifold,'' [arXiv:hep-th/0201028].}

\lref\klst{S. Kachru, X. Liu, M. Schulz and S. P. Trivedi, ``Supersymmetry
Changing Bubbles in String Theory,'' [arXiv:hep-th/0205108].}

\lref\GVW{S. Gukov, C. Vafa and E. Witten,
``CFTs from Calabi-Yau Fourfolds,'' Nucl. Phys. {\bf B584}, 69
(2000) [arXiv:hep-th/9906070].}

\lref\Gukovwal{S. Gukov, ``Solitons, Superpotentials and Calibrations,''
Nucl. Phys. {\bf B574}, 169 (2000) [arXiv:hep-th/9911011].}

\lref\DRS{K. Dasgupta, G. Rajesh and S. Sethi, ``M-theory, Orientifolds
and G-flux,'' JHEP {\bf 9908}, 023 (1999) [arXiv:hep-th/9908088].}

\lref\asym{
J. Harvey, G. Moore and C. Vafa, ``Quasicrystalline Compactification,''
Nucl. Phys. {\bf B304}, 269 (1988)\semi
M. Dine and E. Silverstein, ``New M Theory Backgrounds with Frozen
Moduli,'' [arXiv:hep-th/9712166]\semi
A. Dabholkar and J. Harvey, ``String Islands,'' JHEP {\bf 9902},
006 (1999) [arXiv:hep-th/9809122].}

\lref\Becker{K. Becker and M. Becker, ``M-theory on Eight Manifolds,''
Nucl. Phys. {\bf B477}, 155 (1996) [arXiv:hep-th/9605053].}

\lref\becker{K. Becker and M. Becker, ``Supersymmetry Breaking,
M Theory and Fluxes,'' JHEP {\bf 0107}, 038 (2001) [arXiv:hep-th/0107044].}

\lref\KPV{S. Kachru, J. Pearson and H. Verlinde, ``Brane/Flux Annihilation
and the String Dual of a Non-Supersymmetric Field Theory,''
[arXiv:hep-th/0112197].}

\lref\louis{M. Haack and J. Louis, ``M-theory Compactified on a
Calabi-Yau Fourfold with Background Flux,'' Phys. Lett. {\bf B507},
296 (2001) [arXiv:hep-th/0103068].}

\lref\curio{G. Curio, A. Klemm, D. L\"ust and S. Theisen, ``On the
Vacuum Structure of Type II String Compactifications on Calabi-Yau
Spaces with H Fluxes,'' Nucl. Phys. {\bf B609}, 3 (2001) [arXiv:
hep-th/0012213].}

\lref\EvaFix{E. Silverstein, ``(A)dS Backgrounds from Asymmetric
Orientifolds,'' [arXiv:hep-th/0106209].}

\lref\joefrey{A. Frey and J. Polchinski, ``${\cal N}=3$ Warped
Compactifications,'' [arXiv:hep-th/0201029].}

\lref\GKP{
S.~B.~Giddings, S.~Kachru and J.~Polchinski,
``Hierarchies from fluxes in string compactifications,''
[arXiv:hep-th/0105097].
}
\lref\hull{
C.~M.~Hull,
``Gravitational duality, branes and charges,''
Nucl.\ Phys.\ B {\bf 509}, 216 (1998)
[arXiv:hep-th/9705162].
}

\lref\massive{C. Hull, ``Massive String Theories from M Theory and
F Theory,'' JHEP {\bf 9811}, 027 (1998) [arXiv:hep-th/9811021].}

\lref\GPOne{
M.~Grana and J.~Polchinski,
``Gauge/gravity duals with holomorphic dilaton,''
[arXiv:hep-th/0106014].
}
\lref\GPTwo{
M.~Grana and J.~Polchinski,
``Supersymmetric three-form flux perturbations on AdS(5),''
Phys.\ Rev.\ D {\bf 63}, 026001 (2001)
[arXiv:hep-th/0009211].
}

\lref\Curiodual{G. Curio, A. Klemm, B. Kors and D. L\"ust, ``Fluxes in
Heterotic and Type II String Compactifications,'' [arXiv:hep-th/0106155].}

\lref\andyjoe{J. Polchinski and A. Strominger, ``New Vacua for Type II String
Theory,'' Phys. Lett. {\bf B388}, 736 (1996) [arXiv:hep-th/9510227].}

\lref\mich{J. Michelson, ``Compactification of Type IIB Strings to
Four-Dimensions with Nontrivial Classical Potential,''
Nucl. Phys. {\bf B495}, 127 (1997) [arXiv:hep-th/9610151].}

\lref\dallgata{G. Dall'Agata, ``Type IIB supergravity compactified on
a Calabi-Yau manifold with H-fluxes,'' [arXiv:hep-th/0107264].}

\lref\Moore{
G.~W.~Moore,
``Arithmetic and attractors,''
[arXiv:hep-th/9807087].
}

\lref\Mayr{P. Mayr, ``On Supersymmetry Breaking in String Theory
and its Realization in Brane Worlds,'' Nucl. Phys. {\bf B593}, 99
(2001) [arXiv:hep-th/0003198]\semi
P. Mayr, ``Stringy Brane Worlds and Exponential Hierarchies,''
JHEP {\bf 0011}, 013 (2000) [arXiv:hep-th/0006204].}

\lref\RS{L. Randall and R. Sundrum, ``A Large Mass Hierarchy
from a Small Extra Dimension,'' Phys. Rev. Lett. {\bf 83}, 3370
(1999) [arXiv:hep-ph/9905221].}

\lref\Verlinde{H. Verlinde, ``Holography and Compactification,''
Nucl. Phys. {\bf B580}, 264 (2000) [arXiv:hep-th/9906182]\semi
C. Chan, P. Paul and H. Verlinde, ``A Note on Warped String
Compactification,'' Nucl. Phys. {\bf B581}, 156 (2000)
[arXiv:hep-th/0003236].}

\lref\HassanBV{
S.~F.~Hassan,
``T-duality, space-time spinors and R-R fields in curved backgrounds,''
Nucl.\ Phys.\ B {\bf 568}, 145 (2000)
[arXiv:hep-th/9907152].
}

\lref\GPrefs{
J.~H.~Schwarz and P.~C.~West,
``Symmetries And Transformations Of Chiral N=2 D=10 Supergravity,''
Phys.\ Lett.\ B {\bf 126}, 301 (1983)\semi
J.~H.~Schwarz,
``Covariant Field Equations Of Chiral N=2 D=10 Supergravity,''
Nucl.\ Phys.\ B {\bf 226}, 269 (1983)\semi
P.~S.~Howe and P.~C.~West,
``The Complete N=2, D=10 Supergravity,''
Nucl.\ Phys.\ B {\bf 238}, 181 (1984).
}

\lref\HellermanAX{
S.~Hellerman, J.~McGreevy and B.~Williams,
``Geometric constructions of nongeometric string theories,''
[arXiv:hep-th/0208174].
}

\lref\ghm{
R.~Gregory, J.~A.~Harvey and G.~W.~Moore,
``Unwinding strings and T-duality of Kaluza-Klein and H-monopoles,''
Adv.\ Theor.\ Math.\ Phys.\  {\bf 1}, 283 (1997)
[arXiv:hep-th/9708086].
}

\lref\simeonlattice{
S.~Hellerman,
``Lattice gauge theories have gravitational duals,''
[arXiv:hep-th/0207226].
}

\lref\KaloperYR{
N.~Kaloper and R.~C.~Myers,
``The O(dd) story of massive supergravity,''
JHEP {\bf 9905}, 010 (1999)
[arXiv:hep-th/9901045].
}

\lref\ScherkZR{
J.~Scherk and J.~H.~Schwarz,
``How To Get Masses From Extra Dimensions,''
Nucl.\ Phys.\ B {\bf 153}, 61 (1979).
}

\lref\PapadopoulosNB{
G.~Papadopoulos,
``Global aspects of symmetries in sigma models with torsion,''
[arXiv:hep-th/9406176].
}

\lref\nilmath{
J.~Shin,
``Group Actions on the 3-Dimensional Nilmanifold,''
Trends in Mathematics {\bf 1}, 62 (1998).
}

\lref\BHO{
E.~Bergshoeff, C.~Hull and T.~Ortin,
``Duality in the type-II superstring effective action,''
Nucl.\ Phys.\ B {\bf 451}, 547 (1995).
}

\lref\joeP{
J.~Polchinski,
{\it String Theory, Vol. II}, Cambridge University Press (1998). 
}

\lref\nakahara{
M.~Nakahara,
{\it Geometry, Topology and Physics},
Institute of Physics Publishing (1990). 
}

\lref\TnT{
P.~K.~ Tripathy and S.~P.~Trivedi, 
``Compactifications with Flux on $K3$ and Tori,"
to appear.
}

\lref\ferrara{
S. Ferrara and  M. Porrati,
``N=1 No Scale Supergravity From IIB Orientifolds,''
Phys. Lett. {\bf 545}, 411 (2002)
[arXiv:hep-th/0207135].
}

\lref\dauria{
R. D'Auria, S. Ferrara and S. Vaula,
``N=4 Gauged Supergravity And a IIB Orientifold With Fluxes,''
New J. Phys. {\bf 4}, 71 (2002)
[arXiv:hep-th/0206241].
}

\lref\lledo{
L. Andrianopoli, R. D'Auria, S. Ferrara and  M. A. Lledo,
``Gauging of Flat Groups in Four Dimensional Supergravity,''
JHEP {\bf 0207}, 010 (2002)
[arXiv:hep-th/0203206].
}

\Title{\vbox{\baselineskip12pt
\hbox{hep-th/0211182}
\hbox{CALT-68-2408}
\hbox{SLAC-PUB-9542}
\hbox{SU-ITP-02/38}
\hbox{TIFR/TH/02-32}
}}
{\vbox{\centerline{New Supersymmetric String Compactifications}}}

\centerline{Shamit Kachru $^{a}$\footnote{$^1$}{skachru@stanford.edu},
Michael B. Schulz $^{b}$\footnote{$^2$}{mschulz@theory.caltech.edu},
Prasanta K.  Tripathy $^{c}$\footnote{$^3$}{prasanta@theory.tifr.res.in} and
Sandip P. Trivedi $^{c}$\footnote{$^4$}{sandip@tifr.res.in}}
\medskip\centerline{$^a$ \it Department of Physics and SLAC, 
Stanford University}
\centerline{\it Stanford, CA 94305/94309, USA}
\medskip
\centerline{$^b$ \it California Institute of Technology, 452-48}
\centerline{\it Pasadena, CA 91125, USA}
\medskip
\centerline{$^c$ \it Tata Institute of Fundamental Research}
\centerline{\it Homi Bhabha Road, Mumbai 400 005, INDIA}

\vskip .3in

We describe a new class of supersymmetric string compactifications to
4d Minkowski space.  These solutions involve type II strings
propagating on (orientifolds of) non Calabi-Yau spaces in the presence
of background NS and RR fluxes.  The simplest examples have
descriptions as cosets, generalizing the three-dimensional
nilmanifold. They can also be thought of as twisted tori.
 We derive a formula for the (super)potential governing
the light fields, which is generated by the fluxes and certain
``twists'' in the geometry.  Detailed consideration of an example
also gives strong evidence that in some cases,  
these exotic geometries are related
by smooth transitions to standard Calabi-Yau or $G_2$ compactifications
of M-theory.  

\Date{November 2002}

\newsec{Introduction}

The study of Calabi-Yau compactifications \CHSW\ has been tremendously
fruitful.  These models play an important role in deriving realistic
models of low-energy particle physics from the heterotic string.
Their study has also lead to fascinating discoveries about stringy
geometry and mathematical physics \Essays.

More recently, it has become clear that Calabi-Yau compactifications
of type II strings are also a reasonable starting point for model
building.  In orientifolds of such models, D-branes wrapping various
cycles give rise to non-Abelian gauge groups, and can yield a
reasonable facsimile of the Standard Model \phenom.  Background RR and
NS fluxes can also be included, and generate a computable
(super)potential which stabilizes many of the moduli fields
\refs{\GVW,\DRS,\TayVaf,\curio,\louis,\GKP,\EvaFix,\becker,\kst,\joefrey,\micu,
\ferretti}.  In addition, the fluxes and branes backreact on the bulk
geometry, yielding warped compactifications
\refs{\Becker,\DRS,\Greene}.  In special cases, the warping can be a
large effect \refs{\Verlinde,\GKP,\Mayr,\GidOl}. This provides a
natural mechanism for explaining hierachies of scales, as suggested by
Randall and Sundrum \RS.

However, such Calabi-Yau orientifolds are merely the tip of the
iceberg.  One way to see this is as follows.  Starting from the type
IIB theory on a Calabi-Yau space $M$ in absence of NS three-form flux
${\cal H}$, one can construct a dual type IIA description of the same
physics in terms of a mirror Calabi-Yau $W$.  On the other hand, in
the presence of generic ${\cal H}$-flux on $M$, the mirror geometry
$W$ can no longer be a Calabi-Yau space.  This is clear because mirror
symmetry is a kind of generalization of T-duality \SYZ, and T-duality
maps nontrivial ${\cal H}$ to an interesting deformation of the T-dual
metric.

In this way, starting from a solution of the IIB theory on a
Calabi-Yau orientifold with background fluxes, one can generate mirror
IIA geometries which are not (orientifolds of a) Calabi-Yau, but are
supersymmetric solutions of the equations of motion.  It should be
clear that more generically, one can construct models which are not
described by Calabi-Yau geometries in ${\it either}$ picture.  For
instance, turning on generic NS flux in the IIA non Calabi-Yau
geometry will result in solutions whose IIB dual is no longer a
Calabi-Yau space.

The resulting models are of interest for phenomenological reasons as
well.  They are typically expected to have fewer moduli than
conventional Calabi-Yau compactifications.  In addition, the
computable potential on moduli space could, in some cases, be useful
from a cosmological viewpoint (for a recent investigation in this
direction see \freycos).

Here, we describe the simplest examples of such new supersymmetric
compactifications of type II strings. 	 
Our starting point is the $T^6/Z_2$ orientifold of
type IIB theory in the presence of NS and RR three-form fluxes,
discussed in detail in \refs{\kst,\joefrey}.  In simple enough cases
these models possess classical isometries, and by performing Buscher
duality transformations \buscher, one is able to construct dual
solutions which are not Calabi-Yau geometries.  Instead, we find that
the dual manifolds are well described as cosets, generalizing the
three-dimensional nilmanifold. They can also be thought of as twisted
tori, first studied in \ScherkZR.  The light scalar fields in these
models are governed by a (super)potential which is generated by both
the fluxes and certain gravitational charges which can be formed from
the background metric \hull.

This paper concerns itself with type II strings, but similar questions
are also of interest in the context of the heterotic string.  
Once again, one would like to ask about supersymmetric vacua in
the presence of the three-form NS flux \Andy.  Little is known
about such models, but in some cases these heterotic vacua are
dual to the type II ones. 

The organization of this paper is as follows.  In \S2, we briefly
review the class of models studied in \kst\ and discuss how T-duality
in the presence of ${\cal H}$-flux generates twisted tori by describing a
simple toy example.  In \S3, we derive various dual descriptions of
the $T^6/Z_2$ flux vacua, which exhibit generalizations of the
twisting encountered in \S2.  For illustration we present a detailed
discussion of an ${\cal N}=2$ supersymmetric compactification to 4d
with several dual descriptions involving non Calabi-Yau geometries.
We also find that consideration of the M-theory limit indicates that
this example can be connected (on the moduli space of vacua) to a
smooth type II Calabi-Yau compactification (with no orientifolding).
In \S4, we directly derive the supersymmetry conditions in the dual
pictures, and write down an effective 4d superpotential which imposes
these constraints.  We also verify that this superpotential correctly
reproduces expected domain wall tensions, in the spirit of
\refs{\GVW,\Gukovwal}.  We close with some comments on further
directions for research in \S5.  In two appendices, we summarize the
T-duality formulae for the transformations of IIA/IIB fields and
spinors which are used throughout the paper.

We end by commenting on some recent related  work. 
The papers \refs{\ferrara, \dauria, \lledo}
discuss  flux compactifications and their relation to  
gauged supergravity.   
While this work was in progress, we also became aware of several related
projects.  Work which has significant overlap with the present paper
appeared recently in \refs{\Janetal,\Humbgrp}.  Heterotic string
compactifications on non Calabi-Yau spaces related by duality to the
models studied in \kst\ have appeared in \DasBeck\ (and earlier work
in this direction appeared in \refs{\Andy,\DRS}).  Other
compactifications which involve ``duality twists'' and are related
indirectly to the twisted tori which appear in \S2 have appeared in
\refs{\Atish,\HellermanAX}.  Finally, some of our results were 
also found by S. Gukov \Gukovun.

\newsec{Basic Formalism}

\subsec{Flux vacua in IIB on $T^6/Z_2$}

The theories that we will discuss in subsequent sections are all
related via T-duality to IIB theory compactified on a $T^6/Z_2$
orientifold.  In this orientifold, the $Z_2$ acts to invert all six
circles.  In the absence of flux, the theory is T-dual to the type I
theory, and preserves ${\cal N}=4$ supersymmetry.  Since there are 64
O3 planes located at the fixed-points of the $Z_2$, RR tadpole
cancellation requires that there also be 16 space-filling D3 branes.
These are the T-duals of 16 D9 branes of SO(32) in Type I, and the low
energy effective field theory is the same $SO(32)$ ${\cal N}=4$
supersymmetric Yang-Mills theory, coupled to ${\cal N}=4$
supergravity, familiar from Type I.

However, this is not the most general solution to the $T^6/Z_2$
tadpole cancellation condition.  This class of compactifications also
admits different superselection sectors in which we turn on quantized
NS-NS and RR three-form fluxes.  Let ${\cal H}_3$ denote the NS-NS
flux and $F_3$ denote the RR flux.\foot{The calligraphic font on the
NS-flux indicates that it is a quantity associated with the IIB
orientifold that we are now discussing, and not with one of the T-dual
theories that we will describe in later sections.  Confusion is less
likely for the RR flux, which changes rank under T-duality, so we do
not use the calligraphic notation in the RR-sector.}  These
field-strengths satisfy a Dirac quantization condition
\eqn\quant{{1\over {(2\pi)^2 \alpha^\prime}}\int_\gamma F_3 = m_{\gamma}
\in {\bf Z},~~{1\over {(2\pi)^2 \alpha^\prime}} \int_\gamma {\cal H}_3 
= n_{\gamma} \in {\bf Z},}
where $\gamma$ labels the classes in $H_{3}(T^6,{\bf Z})$. 
%
%
In the presence of such fluxes, the D3 brane charge tadpole
cancellation condition reads \foot{As in \kst, we ignore the
possibility of exotic O3 planes.  To consistently do this, we must
choose the integers in \quant\ to be even, as explained in \joefrey.
This will be sufficient for our purposes.}:
\eqn\tadpole{{1\over 2}N_{\rm flux} + N_{D3} ~=~16}
where
\eqn\defn{N_{\rm flux} ~=~{1\over {(2\pi)^4 \alpha'^2}}\int_{T^6}
{\cal H}_3 \wedge F_3~.}  
So if we turn on fluxes, we should in general introduce fewer D3
branes.  

Nonvanishing fluxes give rise to an effective superpotential for the
Calabi-Yau complex structure moduli \GVW. (A detailed derivation of
this in the context of IIB orientifolds was given in App.~A of \GKP).
The superpotential is
\eqn\superpot{W = \int G_3 \wedge \Omega,}
where
\eqn\Gdef{G_3 = F_3 - \phi {\cal H}_3}
and $\phi$ is the IIB axio-dilaton.  It follows that supersymmetric
vacua are located at points in complex structure moduli space where
$G$ is of type (2,1) and imaginary self-dual.  Furthermore, for
supersymmetry, the K\"ahler structure $J$ should be chosen to make
$G_3$ primitive ({\it i.e.} satisfy $J \wedge G_3 =0$).  These
conditions were studied in detail for the case of $T^6/Z_2$ in \kst,
and it was found that for generic choices of the fluxes \quant\ there
are no supersymmetric critical points.  However, for suitable
non-generic choices of flux, one can find vacua with ${\cal N}=1,2,3$
supersymmetry (and reduced numbers of moduli).

In the absence of flux, these models admit isometries and various
T-dual descriptions exist.  The RR three-form flux transforms quite
simply under T-duality.  Our goal in the following subsections will be
to explore what happens when one tries to construct analogous T-dual
descriptions in the presence of nontrivial ${\cal H}_3$-flux.

Before moving on, we should note one further interesting feature of
these models.  The fluxes and transverse branes and O-planes act as
sources for a nontrivial ${\it warping}$ of the metric.  The 10-metric
in string frame takes the form
\eqn\warpmet{ds^2 = e^{-2A} \eta_{\mu\nu} dx^\mu dx^\nu + 
e^{2A} \tilde g_{mn} dx^m dx^n}
where $\mu,\nu=0,1,2,3$ and $m,n=4,\cdots,9$.  The metric $\tilde
g_{mn}$ is the unwarped metric on the compactification space.  The
warp-factor $e^{2A}$ is determined by the equation
\eqn\warpeq{-\tilde\nabla^2 e^{4A} = (2\pi)^4 (\alpha^\prime)^{2}
g_{s} \tilde \rho_{3} + {g_s \over 12} G_{mnp} \overline
G^{\widetilde{mnp}}}  
where tildes denote the use of the unwarped metric, and $\rho_3$
refers to the localized D3-charge density (which gets contributions
from both D3 branes and O3 planes).  
One can
argue as in \refs{\GKP,\joefrey}\ that under rescaling the metric
$\tilde g_{mn} \to \lambda^{2} \tilde g_{mn}$, the warp factor behaves
like $e^{2A} \sim 1 + O(\lambda^{-4})$.  Therefore, at large radius and
weak coupling, where one trusts the supergravity equations, the
corrections due to the warp factor are negligible.  In the examples we
provide in this paper, one can choose the moduli to lie in a regime
where one can neglect the warping.

\subsec{A warm-up: The twisted torus}

As a preliminary indication of what to expect, let us consider the
following simple toy model.  Imagine starting with a square
three-torus $M$ parametrized by $x,y,z$, with metric
\eqn\threemet{ds^2 = dx^2 + dy^2 + dz^2 .}
Suppose one has also turned on $N$ units of NS three-form flux through
this torus
\eqn\backns{{1\over { (2\pi)^2 (\alpha^\prime)}}\int_{M} {\cal H}_{3}
~=~N .}
We can choose a gauge where
\eqn\bfield{{\cal B}_{yz} =  Nx}
with other components vanishing.  (From now on, we set $(2\pi)^2\alpha'
= 1$ for convenience).

Now, imagine T-dualizing along the $z$ direction.  In the T-dual, the
$B$ field vanishes.  The resulting T-dual metric is

\eqn\dualmet{ds^2 = dx^2 + dy^2 + (dz + Nx dy)^2}
We will call such a space a ``twisted torus'' or a ``nilmanifold,''
following \KaloperYR\ and \nilmath\ respectively.  The identifications
to be made in interpreting \dualmet\ are not the same as in a standard
$T^3$.  Instead one should identify\foot{The nontrivial identification
$(x,y,z)\cong(x+1,y,z-Ny)$ is necessary in order that $dz+Nxdy =
d(z-Ny)+N(x+1)dy$ be globally well-defined.}

\eqn\idents{(x,y,z) \cong (x,y+1,z) \cong (x,y,z+1) \cong (x+1,y,z-Ny).}

This space has another convenient description, as a coset.  Consider
$R^3$, presented as the space of upper triangular $3\times3$ matrices
with ones along the diagonal:

\eqn\rthree{g_N(x,y,z) = \pmatrix{1&y&-{1\over N}z\cr
                     0&1&x\cr
                     0&0&1}}
where $x,y,z$ are real numbers.  Let us call this group ${\cal
G}^N_3({\bf R})$.  For any $N$, this group is isomorphic to the
three-dimensional Heisenberg group ${\cal H}_3$.  We can also define
${\cal G}^N_3({\bf Z})$ by considering analogous matrices

\eqn\rthreet{g_N(a,b,c) = \pmatrix{1&b&-{1\over N}c\cr
                      0&1&a\cr
                      0&0&1}}
with $a,b,c$ integers.  This has a natural action on ${\cal G}^N_3
({\bf R})$ by matrix multiplication.  Consider the right-coset $M =
{\cal G}^N_3({\bf R})/{\cal G}^N_3({\bf Z})$.  The resulting
identifications are
\eqn\result{(x,y,z) \sim (x+a,y+b,z-Nby+c).} 
A little thought shows that the spaces \result\ are the same as the
twisted tori \idents.

Since these spaces were obtained as a {\it right\/}-coset, we
would naively expect the twisted tori to possess an isometry group
corresponding to {\it left\/}-multiplication by ${\cal G}^{N}_{3}({\bf
R})$.  However, the one-forms appearing in the metric \dualmet\ are
the right-invariant forms $\eta^1 = dx$, $\eta^2=dy$, $\eta^3 =
dz+Nxdy$, defined by $g_N(\eta^1,\eta^2,\eta^3) = dg_N^{\vphantom{-1}}
g_N^{-1} (x,y,z)$.\foot{Note that the positive-definite metric
\dualmet\ is not same as the Cartan metric ${\rm Tr}\,dg\,dg$, which
is easily seen to vanish.  (Since the group is non-semi-simple, the
usual theorem that the Cartan metric is positive-definite does not
apply).}  The would-be Killing vectors that generate
left-multiplication are the right-invariant vector fields $k_1 =
\partial_x$, $k_2 = \partial_y - Nx\partial_z$, $k_3 = \partial_z$,
dual to the $\eta^i$.  Of these, only $k_3$ is actually an isometry of
the metric.\foot{The reader is invited to check that the nonvanishing
Lie derivatives of the form ${\cal L}_k\eta$ are ${\cal L}_{k_1}\eta^3
= Ndy$ and ${\cal L}_{k_2}\eta^3 = -Ndx$.  Consequently, only ${\cal
L}_{k_3}$ acting on the metric is nonzero.\par  Another way to understand
this is in terms of the generators of {\it right}-multiplication.  All
of these generators are compatible with the metric, but only $k_3$
(which has a dual interpretation as generator of either left- or
right-multiplication) is compatible with the quotienting that defines
the coset.  That is, $k_3$ is the only generator of
right-multiplication that lies in the commutant of ${\cal G}^N({\bf
Z})$ in ${\cal G}^N({\bf R})$.} The isometry is the $U(1)$ of
translations in the $z$-direction.


It is easy to see that the twisted torus is topologically distinct
from the untwisted torus.  
For instance, 
$h^1(M) = 2$ for $M=$ the twisted $T^3$.  One can prove this as follows. 
Since ${\cal G}^N_3({\bf R})$ is
topologically $R^3$, it follows that $\pi_1\bigl({\cal G}^N_3({\bf
R})/{\cal G}^N_3({\bf Z})\bigl) = {\cal G}^N_3({\bf Z})$.  The
homology group $H_1(M,{\bf Z})$ is the abelianization of this group,
that is, the group ${\cal G}^N_3({\bf Z})$ modulo its commutator
subgroup.  (Here commutator mean group-commutator $XYX^{-1}Y^{-1}$, as
opposed to algebra-commutator $XY-YX$).  It is easy to show that the
commutator subgroup is generated by $e^N$, where $e = g_N(0,0,1)$.
The group $H_1({\bf Z})$ is then ${\bf Z}\times{\bf Z}\times{\bf
Z}_N$, with the first factor generated by $g_N(1,0,0)$, the second by
$g_N(0,1,0)$, and the third by $e$ modulo $e^N$.  The real-valued
homology group is $H_1(M,{\bf R})={\bf R}^2$, of dimension
two.\foot{The interpretation of the torsion factor ${\bf Z}_N$ is that
that there is a third one-cycle, around which string winding is
conserved modulo $N$.  This is the T-dual of the statement that
momentum on the original $T^3$ is conserved modulo $N$ in the presence
of $N$ units of NS-flux.}

Using Cartan's structure equation
\eqn\cartan{d\eta^a + \omega^a{}_b\wedge\eta^b = T^a = 0,
\qquad T^a = \hbox{torsion},}
it is straightforward to solve for the spin connection of the
twisted $T^3$.  We have 
\eqn\niletas{d\eta^1 = d\eta^2=0,\quad d\eta^3 = N\eta^1\wedge\eta^2,}
so the solution is
\eqn\nilspincon{\omega^1{}_2 = -\half N\eta^3,\quad
\omega^2{}_3 = \half N\eta^1,\quad
\omega^3{}_1 = \half N\eta^2.}
Taking the antisymmetric part of the spin-connection, we obtain a
three-form
\eqn\dualH{\omega_{(3)} = \eta^a\wedge\eta^b\wedge\omega_{ab} 
= N \eta^1\wedge\eta^2\wedge\eta^3.}

This illustrates a general rule.  When a component of NS flux is lost
through T-duality, it reappears in the antisymmetrized spin-connection
as 
\eqn\omegak{\omega_{(3)} = k\wedge F_{\rm KK} + \ldots,}
where $k$ is the Killing one-form of the isometry, and $F_{\rm KK}$ is
the flux of the corresponding Kaluza-Klein gauge field \hull.  In this
case, $A_{\rm KK} = Nxdy$, $F_{\rm KK} = Ndx\wedge dy$, and $k =
g_{zm}dx^m = dz+Nxdy$.

This toy model is not really a solution of the string equations of
motion: If one tries to compactify on a $T^3$ with ${\cal H}_3$-flux, both 
the string coupling and the volume of the torus have tadpoles and want to
relax to extreme values.  However, in the more elaborate backgrounds
reviewed in \S2.1, one can find stable toroidal compactifications with
both ${\cal H}_3$ and ${\cal F}$ flux.  Then, dualizing along any isometry 
directions will yield new stable vacua, with ``twistings'' quite analogous 
to the one above.  These vacua will also have a convenient description as
cosets.

One final comment before we move on.  Starting with the metric
\threemet\ and ${\cal B}$ field \bfield, we have discussed the geometry
which results after one T-duality. It is interesting to ask what
happens on doing additional T-dualities. The geometry \dualmet\ (with
$B=0$) is independent of the $y$ direction.  T-dualizing further along
this direction gives rise (using the T-duality rules in App.~A) 
to the metric
\eqn\twotmet{ds^2={1 \over 1+ N^2 x^2}(dz^2+dy^2) + dx^2}
and ${\cal B}$ field
\eqn\twotb{{\cal B}_{yz}={N x \over 1+ N^2 x^2}.}
This background looks quite puzzling at first. For example, on going
around the $x$ circle the metric of the $z,y$ two-torus,
$T^2_{\{yz\}}$, is not periodic even up to an $SL(2, {\bf Z})$ transformation
\twotmet.

A little more thought shows that the metric \twotmet\ and ${\cal B}$ field
\twotb\ are in fact periodic up to an element of $O(2,2,{\bf Z})$ which does
not belong to $SL(2,{\bf Z})$.  This can be understood as follows.  In the
metric \dualmet, on going around the $x$ circle, the two-torus
$T^2_{\{yz\}}$ is twisted by an element of $SL(2,{\bf Z})$ which we denote
as $A$.  T-duality along the $y$ direction, which we denote as $T_y$,
does not commute with $A$. As a result in the final solution the
metric and ${\cal B}$ field, \twotmet\ and \twotb, twist by the transformation
$T_y^{-1} A T_y \in O(2,2,{\bf Z})$ on going around the $x$ circle.

The facts above imply that the supergravity approximation is
not adequate to describe this background. 
Momentum modes along the $T^2_{yz}$ mix with winding
modes on going around the $x$ circle.  Since this paper restricts
itself to the supergravity approximation, we do not consider
backgrounds of this sort any
further.  We follow a simple rule to avoid them: never T-dualize twice
along two directions $a,b$, if $\left[{\cal H}_3\right]_{abc}\ne 0$ 
for any direction $c$ in the starting
configuration.  Twists by elements of the $O(2,2,{\bf Z})$ duality group
have been considered in \refs{\Atish,\HellermanAX}.

\subsec{Twisted Tori in General}


The nilmanifold and related generalizations we discussed above are in
fact examples of twisted tori discussed in the seminal paper of Scherk
and Schwarz \ScherkZR.

One way to think about twisted tori in general is as follows (see also
\KaloperYR): They are parallelizable manifolds with a well defined,
nowhere vanishing basis of vielbein fields. Below, we denote this basis
of vielbein one-forms as $\eta^a$, $a=1,\ldots,n$. The coordinate basis
one-forms $dx^\alpha$ are related to the vielbein $\eta^a$ by
\eqn\ssa{\eta^a = U(x)^a_\alpha dx^\alpha.}
The quantity $U\in GL(n,{\bf R})$ is a matrix which specifies the
twisting.

In particular the twisting matrix $U$ must satisfy an important
property.  The coefficients $f^a_{bc}$, defined by
\eqn\defs{d\eta^a =-{1 \over 2} f^a_{bc}\eta^b\wedge \eta^c,}
must be constant on a twisted torus. Following \ScherkZR, we refer to
these coefficients as structure constants.

Note that as a result the spin connection is also a constant on a
twisted torus. This follows because the spin connection can be
expressed in terms of the structure constants as
\eqn\defspinc{w^c_{ab}={1 \over 2}(f^c_{ab} -\delta_{bs} \delta^{cj} f^s_{aj}- \delta^{cj} \delta_{as}f^s_{bj}),}
(where $\delta^{bs}$ etc denote the Kronecker delta symbol).  Finally,
it is worth mentioning that in the low-energy theory obtained after KK
reduction on a twisted torus, masses for moduli and (non abelian)
gauge couplings can arise. These depend only on the structure
constants, and have no other depndence on the twisting matrix $U$.

Let us see how this general discussion applies to the examples of
\S2.2.  Consider the vielbein \dualmet\ 
\eqn\defexbein{\pmatrix{\eta^1\cr\eta^2\cr\eta^3} =\pmatrix{dx\cr dy
\cr dz+Nxdy}.} 
Note first that the $\eta^a$ are well defined on the twisted torus. In
particular as mentioned in the footnote after \idents, $\eta^3$ twists
by the appropriate $SL(2,{\bf Z})$ transformation in going around the $x$
circle.  The twisting matrix $U$ can be read off from \defexbein, and
is:
\eqn\defu{U= \pmatrix{1&0&0\cr
                      0&1&0\cr
                      0&Nx&1}}
The structure constants then take the form $f^3_{12}=-f^3_{21}=-N$,
with all other components vanishing. They are indeed constant \foot{
Strictly speaking, once the effects of warping are included the coefficients $f_{bc}^a$ are 
no longer constant. In this sense the compactification is a generalisation of the twisted torus
considered in \ScherkZR.}.  Thus
the generalized nilmanifolds of \S2.2 have all the properties of
twisted tori discussed in \ScherkZR. \foot{There is one more condition
on the structure constants in \ScherkZR, $f^a_{ab}=0$.  This too is met
by the example of \S2.2.}

There is another way to connect our discussion of twisted tori above
with nilmanifolds.  The structure constants $f^a_{bc}$ in general
define a (non-compact) Lie algebra.  For the twisted torus of \S2.2
this is the Heisenberg algebra (with $N$ playing the role of
$\hbar$).  The corresponding group is the group of upper triangular
matrices $G$. The nilmanifolds are cosets of exactly this group by
appropriate discrete subgroups.

\newsec{New geometries from $T^6/Z_2$}
\subsec{A detailed example}

For concreteness, it is helpful to describe explicitly a nontrivial
example where one can see the considerations of the previous
subsections come into play.  As our starting point, we take the ${\cal
N}=2$ supersymmetric flux compactification on $T^6/Z_2$ which played
an important role in \klst.  The fluxes are chosen to be: 

\eqn\fflux{F_3 ~=
~2 dx^1 \wedge dx^2 \wedge dy^3 +  2 dy^1 \wedge dy^2 \wedge dy^3}
\eqn\hflux{{\cal H}_3 ~=
~ 2 dx^1 \wedge dx^2 \wedge dx^3 + 2 dy^1 \wedge dy^2 \wedge dx^3}
(Again, here and in the rest of the paper, we set equal to 1 the
factor of $(2\pi)^2 \alpha'$ that should appear on the RHS when
specifying each of the fluxes).  The factors of 2 are inserted to
avoid various subtleties related to ``exotic'' O3 planes which arise
when the fluxes aren't even.  It follows from the equations of \kst\
that this model has a moduli space of ${\cal N}=2$ supersymmetric
vacua.  Along this moduli space, the $T^6$ looks like a $(T^2)^3$. The
three $T^2$'s lie in the $x^iy^i$ directions, with $i=1,2,3$,
respectively. The complex structure moduli $\tau_{1,2,3}$ of the three
two-tori, together with the axio-dilaton $\phi$, satisfy the equations

\eqn\modone{\phi \tau_3 = -1}
\eqn\modtwo{\tau_1 \tau_2 = -1}
There are also K\"ahler moduli which are constrained only by the
primitivity condition.  For simplicity, we work at a point in moduli
space where the metric is diagonal with radii $R_{x^i}, R_{y^i}$.
The constraints on the moduli \modone\ and \modtwo\ can be written
in terms of the radii using $\tau_{i} = i R_{y^i}/R_{x^i}$.

This is a nongeneric situation; for generic choices of flux, the
complex structure and $\phi$ would have been completely frozen.  But
it is convenient to consider such an example, where the fluxes are
quite simple, for several reasons.  One is that then the dual
geometries are also quite simple.  The other, perhaps more important,
reason is that then one can argue that the dual description is at
large radius (and weak coupling), and hence the dual geometry is
meaningful, for an appropriate regime of parameters.

Choose a gauge where
\eqn\bis{{\cal B}_{x^1 x^3} ~ = ~ 2 x^2,~~{\cal B}_{y^1 x^3} ~=~2 y^2~}
are the nonvanishing components of the ${\cal B}_{\mu\nu}$ field on the
internal space.  It is an easy matter to find various dual geometries
now.  We want to avoid the subtleties which arise, as described in
\S2.2, when one attempts to dualize two directions which appear in a
decomposable piece of $H$.  We see that while avoiding this, we can
still safely perform three T-dualities (along the $x^1$, $y^1$ and
$y^3$ directions) in this model.

Let us make one more comment. In the discussion below, to focus on the essential features,
 we neglect the effects of the warp factor \warpmet, \warpeq. As mentioned in section 2, 
this is a good approximation for large volume. 
More complete formulae with the warp factor can be found in section 4, in the discussion of 
supersymmetry. 

\medskip
\noindent{\it{One T-Duality: A IIA Dual}}
\medskip

There are several choices for the order in which one performs the
dualities.  It is clear that dualizing the $y^3$ circle does not yield
any twisting in the geometry of the dual, so we shall save that for
our last transformation.

We will obtain a more interesting result by first dualizing along,
say, the $x^1$ direction.  After performing one T-duality along this
direction, we obtain a model where the metric now has a nontrivial
$2\times2$ block in the $x^1x^3$ coordinates:
\eqn\dualmet{ds^2 = {1\over R_{x^1}^2} (dx^{1} + 2 x^{2} dx^3)^2
+ R_{x^2}^2 (dx^2)^2 + R_{x^3}^2 (dx^3)^2 + \sum_{i=1}^3 R_{y^i}^2
(dy^i)^2.}
In other words, the $x^1, x^2, x^3$ coordinates are sweeping out the
nilmanifold encountered in \S2.2, while the $y^i$ still live on a
square $T^3$.

In this description, one still has a residual $H_3$-field
\eqn\bhere{H_3 ~=~ 2 dy^1 \wedge dy^2 \wedge dx^3, ~~B_{y^1 x^3} ~=~ 2 y^{2}}
along with RR two-form and four-form fluxes
\eqn\twoform{F_{2} = 2 dx^2 \wedge dy^3}
\eqn\fourform{F_{4} = 2 (dx^1 + 2x^2 dx^3) 
\wedge dy^1\wedge dy^2\wedge dy^3.}
The O3 planes have turned into O4 planes wrapping the $x^1$ circle.

The constraints on moduli \modone, \modtwo\ can be easily re-written 
in terms of the new type IIA variables.  They read
\eqn\modonet{\rho_1 \tau_2 = -1}
\eqn\modtwot{\tilde R_{x^1} \phi \tau_3 = -1}
where $\rho_1$ is the volume modulus $\rho_1 = i\tilde R_{x^1}
R_{y^1}$, $\tilde R_{x^1}= 1/R_{x^1}$ is the radius of the T-dualized
circle in the IIA theory, and all other quantities refer to IIA
variables as well.  It is clear that by choosing the point in moduli
space appropriately, one can make this T-dual description the
effective description, as compared to the IIB starting point.

Notice that it follows from App.~A that the piece of the $H$-field
which had a leg along the $x^1$ direction is encoded after the duality
transformation in the spin-connection $\omega$ in the IIA theory.  In
the notation of App.~A, one finds that
\eqn\newome{g_{(x^1)} = 2 x^2 dx^3,\qquad
\omega_{(x^1)} = -2 dx^2\wedge dx^3.} 
Here, $g_{(x^1)}$ is the Kaluza-Klein gauge-field corresponding to the
isometry in the $x^1$-direction, and $-\omega_{(x^1)}=dg_{(x^1)}$ is
the corresponding field-strength.  The Killing one-form corresponding
to this isometry is $k= \eta^1/R_{x^1}^2$, where $\eta^1 =
(dx^1+2x^2dx^3)$.  So, as in \S2.2 and \hull, the antisymmetrized
spin-connection is
\eqn\newantispin{\omega_{(3)} = k\wedge F_{(2)}^{\rm KK} 
= {1\over R_{x^1}^2} dx^1\wedge dx^2\wedge dx^3.}
Therefore, explaining the moduli constraints \modonet\ and \modtwot\
directly in the IIA theory should require us to write down a
superpotential which has nontrivial dependence on the spin-connection,
or equivalently, on the corresponding Kaluza-Klein flux.  This
expectation will be borne out in \S4, where we derive the IIA
superpotential and show that it can correctly reproduce the
constraints \modonet\ and \modtwot\ in this example.

Finally, we note here that this compactification manifold is
non-K\"ahler.  One way to prove this is as follows.  As in \S2.2,
we can use the relationship between the fundamental group and the
first homology of the manifold to compute $h^{1}$.  In this case,
$h^{1}=5$, while for a K\"ahler manifold $h^1$ is  
always even.  As we will describe in \S4.1, this particular example
${\it does}$ admit an integrable complex structure, so it is
a complex but non-K\"ahler space.

\medskip
\noindent{\it{Two T-Dualities: A IIB Dual}}
\medskip

Again, dualizing the $y^3$ coordinate at this stage would not lead to
any further interesting twists in the metric.  So instead we dualize
the $y^1$ coordinate.  The resulting IIB geometry is characterized by
the metric 
\eqn\twodmet{\eqalign{ds^2 &= \tilde R_{x^1}^2 (dx^1 + 2x^2 dx^3)^2 
+ R_{x^2} (dx^2)^2 + R_{x^3} (dx^3)^2\cr
&\qquad\qquad+ {1\over R_{y^1}^2} (dy^1 + 2 y^2 dx^3)^2 +
R_{y^2}^2 (dy^2)^2 + R_{y^3} (dy^3)^2.}}
At this stage of the duality, we have reached a state where the
$B$-field on the internal space vanishes. However, there is a
nontrivial RR three-form flux
\eqn\brrflux{F_{3} = 2 (dx^1 + 2x^2dx^3)\wedge dy^2\wedge dy^3 
+ 2 (dy^1 + 2y^2dx^3)\wedge dx^2\wedge dy^3}

It is simple to T-dualize the moduli constraints \modone\ and \modtwo\
and one again finds that in a suitable regime of moduli space, this 
twice T-dualized geometry is the most effective description of the physics.
One can again quickly argue that this space is non-K\"ahler, by writing
out the fundamental form $J$ which comes from the metric \twodmet.  
This form has $dJ \neq 0$.  
A more complete discussion of this is given in \S4.2, where we
describe the supersymmetry conditions after two T-dualities.

\medskip
\noindent{\it{Three T-Dualities: The IIA ``Mirror''}}
\medskip

Finally, one can perform a third T-duality along the $y^3$ direction.
One might call the result a ``mirror" geometry to our starting point,
since mirror symmetry can be understood as T-duality along
supersymmetric $T^3$ fibers \SYZ.  The effect of this last duality on
the geometry is simply to take the metric from the previous step and
flip $R_{y^3} \to 1/R_{y^3}$.  There is in addition a nonvanishing
$F_2$ flux
\eqn\afrrflux{F_{2} = 2 (dx^1+2x^2 dx^3)\wedge dy^2 
+ 2(dy^1+2y^2dx^3)\wedge dx^2.}

Although this is a mirror description of the original IIB Calabi-Yau
orientifold, the resulting metric is not related to a Calabi-Yau
metric.  The new information is encoded in windings:

\noindent
$\bullet$
The $x^1x^3$ $T^2$ undergoes an SL(2,${\bf Z}$) monodromy
as one goes around the $x^2$ circle.

\noindent
$\bullet$
The $y^1x^3$ $T^2$ undergoes an SL(2,${\bf Z}$) monodromy as one goes
around the $y^2$ direction.

\noindent

Altogether, the nontrivial structure of the metric can be encoded in
some SL(3,${\bf Z}$) matrices that act on the $x^1y^1x^3$ directions, as
one moves around on the base parametrized by $x^2y^2y^3$.  It would
be very interesting to find such non Calabi-Yau mirrors of
(orientifolds of) more generic Calabi-Yau manifolds with flux.
Note that in the example above we T-dualised along the $x^1,y^1,y^3$ direction,
so this is the $T^3$ fibre of mirror symmetry. 
The $SL(3,{\bf Z})$ twist then mixes the fibre and base directions 
of the Calabi-Yau manifold.

After this third T-duality, the moduli constraints become:
\eqn\modonef{\tilde R_{y^1} R_{x^2} = \tilde R_{x^1} R_{y^2}}
\eqn\modtwof{\tilde R_{y^1} \tilde R_{x^1} = g_{s} R_{x^3}}
where now all variables appearing are the appropriate variables
for the final IIA description (each $\tilde R$ is the inverse
of the original starting IIA radius, and $g_s$ is the final IIA
coupling). 
By choosing an appropriate regime of couplings and radii, one can
satisfy these constraints while making the description after three
T-dualities the most effective description.

\subsec{M-theory limit and the web of vacua}
 
Our final IIA background has the following property.\foot{We thank
P. Berglund and N. Warner for emphasizing this point to us.}  By
taking the strong-coupling limit of this IIA vacuum one can get an
M-theory description where all radii (including that of the 11th
dimension) are large in 11d Planck units, consistent with the
constraints \modonef\ and \modtwof.  But since the IIA background was
characterized by only a nontrivial metric and $F_2$, the M-theory
description should be ${\it purely~ geometrical}$.

To obtain 4d ${\cal N}=2$ supersymmetry from M-theory, one should
compactify it on a Calabi-Yau threefold times a circle.  Therefore,
the 11d limit of this model should yield M-theory on some Calabi-Yau
threefold $X$ times a circle.  From this viewpoint, we have done a
twisted reduction (presumably using a twisted circle in the
$T^3$-fibers of the Calabi-Yau threefold) to get our IIA model.  It
may be possible to characterize this threefold more concretely, using
facts about the M-theory lift of the D6 branes and O6 planes.  Since
it is known how to write the M-theory lift of O6 planes in terms of
the Atiyah-Hitchin space, and how to relate D6 branes to ALE metrics
in M-theory \Sen, the construction of the Calabi-Yau $X$ should
involve gluing these local features together into a compact geometry.
In fact, one can obtain a family of models with different numbers of
D6 branes (and different twisting) by rescaling the original fluxes.
In particular, it is possible to rescale the fluxes so that one
satisfies the RR tadpole conditions with {\it no} D6 branes. The
M-theory lift of this model would only involve a gluing of
Atiyah-Hitchin spaces (and the right twisting of the $S^1$ to give
rise to $F_2$) in the M-theory limit.  It would be interesting to more
completely characterize the Calabi-Yau manifolds one gets by lifting
this set of models.

Even without detailed knowledge of the structure of $X$, this result
is interesting because it implies that the orientifold example we have
described here is connected, on the moduli space of vacua, to type IIA
compactifications on smooth Calabi-Yau threefolds (with no
orientifolding).  It was already shown in \klst\ that the model with
{\it no} fluxes, which is just dual to the standard ${\cal N}=4$
supersymmetric compactification of the heterotic string on $T^6$, is
connected by vacuum bubbles (of very low tension, in appropriate
regimes) to the other flux vacua on $T^6/Z_2$.  In particular, our
${\cal N}=2$ model lies on the same configuration space as the
heterotic string on $T^6$ (or type II on $K3 \times T^2$), in a
meaningful sense.  But this, combined with our present result, implies
that the web of ${\cal N}=2$ type II compactifications on Calabi-Yau
spaces is connected (slightly ``off'' the moduli space) to the ${\cal
N}=4$ models, and to the ${\cal N}=1,2$ flux models of \kst.

Finally, it should be clear that we have considered this ${\cal N}=2$
model for its simplicity, and because we are confident that vacua with
extended supersymmetry exist as solutions to the full theory.  However,
our considerations could be repeated with many ${\cal N}=1$ flux vacua
(in the approximation that the flux-generated potential is the full
potential).  In suitable examples, where there is a IIA dual picture
involving only $F_2$ flux, one would then expect to find a 
connection between these flux models and M-theory compactifications on
manifolds of $G_2$ holonomy.

\subsec{The example as a coset}

The example described above can also be characterized simply as a
coset, in analogy with the nilmanifold of \S2.2.  Although we will
discuss the procedure which allows us to do this in our example, it
should be clear that the ideas would generalize to other ``twisted
tori.''  For simplicity, we will provide the coset description of the
5-manifold $Y$ spanned by $(x^1,x^2,x^3,y^1,y^2)$ in \twodmet\ (the
$y^3$ direction is just an extra circle in the geometry), with all
radii set to 1.

A convenient basis of invariant one-forms on 
$Y$ is provided by

\eqn\forms{\eqalign{&\eta^1 = dx^1+2x^2 dx^3,\quad\eta^2 = dx^2,
\quad\eta^3 = dx^3,\cr &\eta^4 = dy^1 +2y^2dx^3,\quad\eta^5 = dy^2.}}
These satisfy an equation of the form 
\eqn\wprod{d\eta^i =  - {1\over 2}f^{i}_{jk}\eta^j \wedge \eta^k,}
where $f$ is antisymmetric on its two lower indices.  The only 
non-vanishing components of $f$ in this case are
\eqn\fcomps{f^{1}_{32} =  - f^{1}_{23} = 2,\quad
f^{4}_{35} = - f^{4}_{53} = 2.} 

Given such ``structure constants'' $f^{i}_{jk}$, we can naturally write
down a Lie algebra with generators $E_i$, satisfying the commutation 
relations: 

\eqn\liealg{[ E_j, E_k ] ~=~ f^{i}_{jk} E_{i}~.} 
A simple matrix representation of generators satisfying these relations
can be written down as follows.  Define the matrices ${\cal E}_{\alpha
\beta}$ which have only one nonzero entry:
\eqn\caledef{({\cal E}_{\alpha\beta})_{ij} ~=~\delta_{\alpha,i} 
\delta_{\beta,j}~.} 
Then a representation of the algebra \liealg\ with structure constants
\fcomps\ is given by choosing
\eqn\gencho{E_{1} = 2{\cal E}_{24},~E_{2} = 2{\cal E}_{23},~
E_{3} = - 2{\cal E}_{34},~E_{4} = 2{\cal E}_{14},~E_{5} = 2{\cal E}_{13}~.} 

It is easy to check that these generators have the property that
at the quadratic level, all products vanish except
\eqn\quadprod{E_{2} E_{3} = - 2E_{1},~~E_{5} E_{3} = - 2E_{4}~.} 
Therefore, when we exponentiate the Lie algebra to form a group $G$, we
find that one can write the generic group element as
\eqn\groupel{g = {\bf 1} + x_1 E_1 + x_2  E_2 + x_3 E_3 + y_1 E_4 + 
y_2 E_5}
Here ${\bf 1}$ is the unit matrix and the $x$s and $y$s are real
numbers. 
We can also define a group $G({\bf Z})$, 
as the group whose elements can
be written as 
\eqn\groupelz{h = {\bf 1} + \sum_{i=1}^{5} n_i E_i} 
with $n_i \in {\bf Z}$. 

Now, consider identifying points in $G$ under left-multiplication by
elements of $G({\bf Z})$.  One sees that
\eqn\lmult{\eqalign{&
({\bf 1} + n_i E_i) ({\bf 1} + x_1 E_1 + x_2 E_2 + x_3 E_3 + y_1 E_4 +
y_2 E_5) = \cr &{\bf 1} + E_1 (n_1 + x_1 - 2n_2 x_3) + E_2 (n_2 + x_2)
+ E_3 (n_3 + x_3) + E_4 (n_4 + y_1 - 2n_5 x_3) + E_5 (n_5 + y_2).}}
Therefore, the identifications generated by the $G({\bf Z})$ action
are precisely those which are characteristic of the twisted geometry
with metric  
\twodmet.
In particular, the twisted identifications
\eqn\twistid{(x^1,x^2,x^3,y^1,y^2) \sim (x^1 - 2x^3,x^2+1,x^3,y^1,y^2) \sim
(x^1,x^2,x^3,y^1 - 2x^3,y^2+1)}
are reproduced by \lmult. 

The structure described here generalizes to the full set of T-duals
of flux vacua on $T^6/Z_2$. 
The invariant one-forms in the twisted picture always define structure
constants via an equation of the form \wprod.  
These structure constants can be used to define a Lie algebra
and a corresponding Lie group $G$, which
admits a simple representation in terms of upper triangular matrices.
The resulting generalized
nilmanifold geometry is a coset of $G$ by the appropriate discrete
subgroup.

\newsec{Supersymmetry}

In this section we analyse the spinor conditions and determine the
requirements imposed by ${\cal N}=1$ supersymmetry in the various
T-dual descriptions.  Starting from the IIB backgrounds in the class
described in \S2.1, we characterize the theories obtained after one
T-duality in \S4.1\ and those obtained after two T-dualities in \S4.2.
The supersymmetry requirements will also fix the superpotential (the
analogue of \superpot) in the dual theories, as is discussed in \S4.3.

Our approach is straightforward.  In the starting IIB theory the
neccessary and sufficient conditions for susy in the presence of flux
are well know \refs{\GPOne,\becker}, and were summarized in \S2.1. We
use the known rules of T-duality \refs{\BHO,\HassanBV}, summarized in
Apps.~A and B, to map these conditions to the dual theories.  This
yields the required conditions for susy in the dual variables.

\subsec{IIA with $O4$ Planes}

We start with the IIB theory with $O3$ planes.  In the discussion
below, we will follow the notation of \GPOne\ closely, see also \kst.
We consider the conditions imposed by supersymmetry on the metric and
flux in the original IIB theory, and map them to conditions in the IIA
theory.

\medskip
\noindent{\it The supersymmetry conditions:}
\medskip

Our notation is as follows.  We denote the direction along which
T-duality is carried out as $x$.  $\mu,\nu$ refer to coordinate
indices along all six internal directions, while $\alpha,\beta$ refer
to coordinate indices along the five internal directions, not
including $x$. Finally, $a,b$ denote the six tangent space indices.
One more comment before we get started. Unlike much of the rest of this paper,
we include the warp factor in this subsection. 

Supersymmetry requires that metric of the starting IIB theory is
conformally $CY_3$ (more generally it is a manifold with reduced
holonomy, in our case $T^6$).\foot{Here we only consider
compactifications where the dilaton is a constant.}  Up to the warp
factor it is therefore complex and K\"ahler.  Let
\eqn\defspb{{\tilde \epsilon} = {\tilde \epsilon}_L + i{\tilde  \epsilon}_R}
denote the spinor in the IIB theory which meets the requirements of
supersymmetry.  

As discussed in \refs{\becker,\GPOne}, $\tilde \epsilon$ is of B-type 
and therefore has definite
$SO(6)$ chirality. 
An almost complex structure (ACS) can then  be defined by 
\eqn\defcb{J^a_b={\tilde \epsilon}^\dagger 
[\gamma^a, \gamma_b] { \tilde \epsilon}}
(we take ${\tilde \epsilon}$ to be appropriately normalised so that $J^2=-1$).
For $T^6/Z_2$ the spinor ${\tilde \epsilon}$ is a constant independent of 
the internal
coordinates, and it is easy to see that the Nijenhuis tensor 
vanishes. As a result the ACS
\defcb\ is integrable.

Now we turn to the metric in the IIA theory obtained by one T-duality. 
We denote the direction along which the T-duality is carried out as $x$. 
We start  with the metric on the $T^6/Z_2$ in IIB
\eqn\startmetb{ds^2= e^{2A} j_{\mu\nu}dx^\mu dx^\nu,}
where $j_{\mu \nu }$ is constant and $A$ is the warp factor.  There is
also a ${\cal B}$ field: \foot{Other components of the $\cal B$ field
do not enter in determining the dual metric.}
\eqn\startb{\left[{\cal B}_2\right]_{x\alpha}
= \left[{\cal H}_3\right]_{x\alpha\beta}x^\beta.}
We find after using the rules for T-duality, App.~A, that the resulting 
metric in IIA can be written as 
\eqn\startmeta{ds^2= e^{-2A}{1\over j_{xx}} \eta^x\eta^x 
+ e^{2A} \left(j_{\alpha \beta} - {j_{\alpha x} j_{\beta x} \over
j_{xx}}\right) \eta^\alpha \eta^\beta.}  
Here, $\bigl\{\eta^x,\eta^\alpha\bigr\}$, $\alpha = 2,\ldots,6$, is a basis of
one-forms defined by
\eqn\defetaspin{ \eta^x= dx - {\cal B}_{x \alpha } dx^\alpha }
and
\eqn\defetaspinb{\eta^\alpha=dx^\alpha,\quad \alpha=2,\ldots,6.}
Using ${\cal B}_{x\alpha} = -g_{x\alpha}/g_{xx}$ (from App.~A), we can
equivalently write \defetaspin\ as
\eqn\killingeta{g_{xx}\eta^x = k_{(x)}\equiv g_{x\mu}dx^\mu.}
The quantitity $k_{(x)}$ in Eq.~\killingeta\ is the Killing one-form.
It is one-form that is dual to the Killing vector $\partial_x$, which
generates translations in the $x$-direction. (In components, $k_{(x)}^\mu =
\delta^\mu_x$, so $k_{(x)\mu} = g_{x\mu}$).

A nowhere vanishing, globally well-defined vielbein field, ${\hat
\eta}^{a}$, $a = 1,\ldots,6$, can be defined in terms of the $\eta^x,
\eta^\alpha$ by diagonalizing the metric \startmeta.  This is in
accord with \S2.3\ where we had discussed that twisted tori are in
general parallelizable manifolds.

Next we turn to the spinor analysis.  The spinor transformations rules
under T-duality show, generally, that if ${\tilde \epsilon}$
\defspb\ meets the spinor conditions for susy in IIB, then the spinor
\eqn\defspa{\epsilon=\epsilon_L+i  \epsilon_R}
meets the required conditions in the IIA theory. Here $\epsilon$ is
defined in terms of the IIB spinor as follows:
\eqn\spba{ {\tilde \epsilon_L} = \Omega  \epsilon_L, \ {\tilde \epsilon_R}=  \epsilon_R.}
$\Omega$ is defined in App.~B.

Consider now the following definition of an ACS in the IIA case.
It is given by the same matrix $J^a_b$ defined in
\defcb, but with the indices $a,b$ taking values in the ${\hat \eta}^a$ tangent
space basis (defined after \defetaspinb) of the dual manifold.
Expressed in terms of the IIA spniors the almost complex structure  takes the form:
\eqn\defacsa{J^a_b={\left(\Omega \epsilon_L 
+ i \epsilon_R\right)}^{\dagger} \left[\gamma^{a},\gamma_{b}\right]
\left(\Omega \epsilon_L + i \epsilon_R\right).}

It is useful to understand this ACS in the basis of one-forms defined in \defetaspin,
\defetaspinb. 
Group the $\eta^x, \eta^\alpha$ into two sets, $\eta^1=\eta^x,\eta^2,\eta^3$
and $\eta^4,\eta^5,\eta^6$. 
Then define holomorphic one forms as 
\eqn\holoeta{\eta^i_{H}=\eta^i+\tau^i_j\eta^{3+j}, ~i,j=1, \cdots 3.}
This choice of holomorphic one forms defines an almost complex
structure (ACS) parametrized by the matrix $\tau^i_j$.  The reader
will notice that the above steps are analogous to parametrizing the
complex structure for a torus.  Keeping that analogy in mind we will
refer to $\tau^i_j$ as the period matrix of the twisted torus
below. Since the spinor $\tilde \epsilon$ \defspb\ is a constant, the
ACS defined in \defacsa\ is a member of this family, and corresponds
to a particular choice of the matrix $\tau^i_j$.

In the basis of holomorphic one-forms the metric \startmeta\ can be
written as
\eqn\metaherm{ds^2= g_{i\jbar} \eta^i_{H} {\bar \eta_{H}}^\jbar,}
where $g_{i\jbar}$ can be easily obtained in terms of the IIA
metric \startmeta. This shows that the metric is Hermitian with respect to the ACS
\defacsa. We can now  define a two-form, called the
fundamental two-form,
\eqn\funda{J_{IIA}= i g_{i \jbar} \eta_H^i \wedge {\bar \eta_H}^\jbar.}

We will see towards the end of this section that the ACS defined above
is not integrable.  One can also show that the fundamental
two-form, \funda, is not closed \foot{For example, one term in $dJ$ arises from,
 $d\eta_H^1=d\eta^x \ne 0$.
Another term arises  because  $g_{i\jbar}$ depends on the warp factor, $A$, \startmetb.}.
 This implies that the
dual geometry is not K\"ahler. We will see below that an integrable ACS can be defined in 
the IIA theory, but even with this choice, the  fundamental two-form 
is not closed and the manifold is not K\"ahler.  This is in agreement with our
discussion of the specific example in \S3.1, where we showed that the first homology
group 
of the dual manifold is odd dimensional.  

The main motivation for considering the ACS above is that the
resulting susy conditions on the flux are easy to state, as we shall
now see. Consider first the requirements imposed by susy on the flux
in the IIB theory.  These are that the $G_3$ flux should be of type
$(2,1)$ and primitive, as reviewed in \S2.1. We now map these
conditions to the IIA side.

As discussed in App.~B, the flux $G_3$ maps on the IIA side to the
three-form
\eqn\gintwofinal{ G_{\rm IIA} = \left( \tilde{F}_{4(x)}
+ g_{xx} \eta^x\wedge F_2 \right) - i \bigl(\sqrt{g_{xx}}/g_s^{\rm
IIA}\bigr) \left( H_3 - g_{xx} \eta^x \wedge \omega_{(x)} \right).}
Eq.~\gintwofinal\ is based notation that is also introduced in App.~B,
we will comment more on the various terms that appear in it  below.

With the choice of ACS specified above in \defacsa\ and \defspa, it then follows, as
we illustrate in App.~B, that the $(2,1)$ and primitivity conditions
map to two requirements on the IIA side.  First, that $G_{\rm IIA}$ is
of type $(2,1)$ with respect to the ACS \defacsa, \defspa.  And second,
that it satisfies the condition:

\eqn\prima{J_{\rm IIA} \wedge G_{\rm IIA}=0,}
where $J_{\rm IIA}$ is the fundamental two-form \funda\ in the IIA
theory.  We will refer to \prima\ as the primitivity condition, since
it is analogous to the corresponding condition in the IIB theory.

Before going further let us discuss the various terms in \gintwofinal\ in more detail.
These arise from the flux $G_3$, \Gdef,   after T-duality. 
${\tilde F}_{4(x)}$ in \gintwofinal, stands for the gauge invariant four-form in IIA
with one leg along the $x$ direction. 
It arises from  $F_{3}$ in the IIB theory with no leg in the $x$ direction.
Similarly, $F_2$ and $H_3$ denote the two form and NS three form in IIA, these terms arise
from $F_3$ with a leg along the $x$ direction and from ${\cal H}_3$  without a leg along the 
$x$ direction respectively. The quantity
$\sqrt{g_{xx}}/g_s^{\rm IIA}$ is simply the dilaton in the IIB theory.
Finally, the last term in \gintwofinal,  arise by T-duality from ${\cal H}_3$ 
with a leg along the $x$ 
direction.  The quantity $k_{(x)}=g_{xx}\eta^x$ is the Killing
one-form \killingeta\ associated with the isometry in the
$x$-direction, and $-\omega_{(x)}$ is the corresponding Kaluza-Klein
field-strength.  Together, $-g_{xx}\eta^x\wedge\omega_{(x)}$ is the
antisymmetrised spin-connection three-form $\omega_{(3)}$,\foot{All other
components of the antisymmetrized spin-connection are required to
vanish by the orientifold projection.} in accord with our discussion in section 2, \omegak.


It is also worth commenting that from a IIA perspective, forgetting about the original 
IIB description,
the direction $x$ is singled out in \gintwofinal\ (and in the IIA
superpotential) because the $O4$ planes extend along it.

Below we comment in more detail on the ACS defined above in the susy analysis. 
In \S4.3, we discuss how to formulate a superpotential in the IIA theory which implements the 
$(2,1)$ condition on the flux, $G_{IIA}$.

\medskip
\noindent
{\it The ACS in more detail}
\medskip

It is clear that the ACS defined in the IIA theory above, \defacsa,
bears a close relation to the complex structure in the starting IIB
theory, \defcb.  However, in the IIA theory, due to the non-trivially
twisted torus, the ACS is not integrable, in general.  The condition
for integrability \nakahara\ is easy to state in terms of the
structure constants which specify the twist, \defs. In the holomorphic
basis \holoeta\ (with $i,\ibar$ denoting holomorphic and
antiholomorphic indices respectively), it takes the form:
\eqn\condint{f^i_{\jbar{\bar k}}=f^\ibar_{jk}=0.}

The lack of integrability is perhaps best illustrated with an example.
Consider the example of one T-duality discussed in \S3.1. We start in
the IIB theory at a point in moduli space where the $T^6$ is of the
form $(T^2)^3$, with the modular parameters of the three $T^2$s being
purely imaginary, of the form $\tau_i=i R_{y^i}/R_{x^i}$.  The
three holomorphic differentials then are $dz^i=dx^i+\tau_i dy^i~.$ For
simplicity we also take the metric to be of form: $ds^2=dz^id{\bar
z}^\ibar$.

In the IIA dual theory the resulting ACS \defacsa\ corresponds to the
choice of the period matrix $\tau^i_j$, \holoeta, of the form
\eqn\choicetau{\tau^i_j={\rm diag}(\tau_1,\tau_2,\tau_3).}
The three holomorphic one forms are then given by 
\eqn\holoexample{\eta^1_H=\eta^x+ \tau_1 \eta^4,\quad
\eta^2_H=\eta^2 + \tau_2 \eta^5,\quad
\eta^3_H=\eta^3+\tau_3 \eta^6.}
We will skip some of the details here, but this can be established by
using the matrix $Q^\mu_\nu$ defined in App.~B, to relate the vielbein
${\hat \eta}^a$, which appears in \defacsa, to the basis \defetaspin,
\defetaspinb, which in turn is related to the holomorphic one forms in
\holoeta.
 
To test for integrability, we consider
\eqn\testi{d\eta^1_H=d \eta^x=\eta^2\wedge \eta^3.}
Expressing the RHS in terms of the holomorphic basis \holoexample,
one sees that $f^1_{{\bar 2} {\bar 3}} \ne 0$, so \condint\ is not met \foot{
We have neglected the effects of warping here. Incorporating them 
does not change the conclusions.
 $\eta_H^1$ is now given by $\eta^1_H=e^{-A}\eta^x+\tau_1e^{A} \eta^4$,
so that  $d\eta^1_H$ has an additional term.}.

We have not explored in general whether an integrable ACS can be
defined in the dual model.  For the example considered above the
answer turns out to be yes.  E.g., consider an alternate choice of
holomorphic one forms
\eqn\altholo{{\eta_H^1=\eta^1+\tau_1\eta^4,\quad
\eta_H^2=\eta^2+ \tau_2 \eta^3,\quad 
\eta_H^3=\eta^5+\tau_3 \eta^6}~.}
Then one can show that $d\eta_H^1\sim \eta_H^2 \wedge {\bar \eta}_H^2$, so 
\condint\ is now met.

\subsec{IIB with $O5$ Planes}

\def\ah{\hat\alpha}
\def\bh{\hat\beta}
\def\gh{\hat\gamma}
\def\detjht{{\rm det}_{xy}{\hat j}}

We turn next to the model obtained by two T-dualities. Since the
analysis is very similar to the one T-duality case, we will be brief
below \foot{We neglect the warp factor here,
this can be included along the lines of  the previous section.}. The two directions along which T-duality is done are denoted by
$x,y$. Using the rules in App.~A the metric after two T-dualities can
be written as
\eqn\twotdual{
ds^2 = {\hat j}_{xx} \eta^x \eta^x + {\hat j}_{yy} \eta^y \eta^y
       + 2 {\hat j}_{xy} \eta^x \eta^y 
       + {\hat j}_{\ah\bh} \eta^{\ah} \eta^{\bh} ~.
}
Here $\eta^x,\eta^y,\eta^{\hat \alpha}$ denote a basis of one-forms given by:
\eqn\etax{\eqalign{
\eta^x & = dx - {\cal B}_{x\ah} dx^{\ah} \cr
& = dx + {{\hat j}_{yy}\over \detjht} \left( {\hat j}_{xx} {\hat j}_{(x)}
       - {\hat j}_{xy} {\hat j}_{(y)} \right),}}
\eqn\etay{\eqalign{
\eta^y & = dy - {\cal B}_{y\ah} dx^{\ah} \cr
& = dy + {{\hat j}_{xx}\over \detjht} \left( {\hat j}_{yy} {\hat j}_{(y)}
       - {\hat j}_{xy} {\hat j}_{(x)} \right),}}
and
\eqn\etahat{\eta^{\ah} = dx^{\ah}.} 
The notation used above is as follows.  The metric components above,
which are denoted with a hat superscript, are constant.
We have also used the definitions $\detjht \equiv
({\hat j}_{xx} {\hat j}_{yy} - {\hat j}_{xy}^2) $ and ${\hat j}_{(x)}
\equiv {\hat j}_{x\ah}dx^{\ah}/{\hat j}_{xx}~,~ {\hat j}_{(y)} \equiv
{\hat j}_{y\ah}dx^{\ah}/{\hat j}_{yy}$.  Finally, the indices
$\ah,\bh$ run over all compact direction except $x,y$.

The analog of Eq.~\killingeta\ is
\eqn\etakillings{\eqalign{%
\hat j_{xx}\eta^x +\hat j_{xy}\eta^y
&= \hat k_{(x)} \equiv \hat j_{x\mu}dx^\mu ,\cr
\hat j_{yx}\eta^x +\hat j_{yy}\eta^y
&= \hat k_{(y)} \equiv \hat j_{y\mu}dx^\mu.}}
Here, $\hat k_{(x)}$ and $\hat k_{(y)}$ are the Killing one-forms that
are dual to the Killing vectors $\partial_x$ and $\partial_y$, which
generate translations in the $x$- and $y$-directions, respectively.
(The determinants in the definition of $\eta^x$,$\eta^y$ cancel out in
the linear combinations on the LHS of \etakillings, leaving the RHS).
 
Once again we define an ACS in the T-dual theory, in terms of the
spinor ${\hat{\tilde\epsilon}}$.  In a vielbein basis ${\hat \eta^{a}}$ which
is obtained by diagonalising the metric above, it takes the form
\defcb. Holomorphic one forms analogous to \holoeta\ can now be
defined, and the metric \twotdual\ can be written in the form
\metaherm\ in this basis.  This allows us to define a fundamental
two-form.  It takes the form:
\eqn\fundtwo{{\hat J}=i~{\hat j}_{i\jbar}\eta^i_H\wedge \eta^\jbar_H.}

As discussed in App.~B, the $G_3$ flux of the starting IIB theory now maps to a three form
\eqn\ibtdul{\eqalign{
{\hat G}_3 = & \left\{ {\hat{\tilde F}}_{5(yx)} + \left( {\hat j}_{xx} \eta^x 
          + {\hat j}_{xy} \eta^y\right) \wedge {\hat{\tilde F}}_{3(y)}
          - \left( {\hat j}_{xy} \eta^x 
          + {\hat j}_{yy} \eta^y\right) \wedge {\hat{\tilde F}}_{3(x)} 
+ \left(\detjht\right) \eta^x \wedge \eta^y \wedge {\hat F}_1 \right\}\cr
& - \bigl(i/{\hat g}_s^{\rm IIB}\bigr) \sqrt{\detjht} \left\{ {\hat{\cal H}}_3 
+ {\hat j}_{xx} \eta^x \wedge d{\hat j}_{(x)}
+ {\hat j}_{yy} \eta^y \wedge d{\hat j}_{(y)}
\right\}.
}}
Our notation is defined in App.~B.
Supersymmetry then leads to the conditions that ${\hat G}_3$ is of
type $(2,1)$ (with reference to the ACS defined above) and primitive
(with respect to the fundamental form \fundtwo).

Let us comment briefly on the various terms in \ibtdul. These arise from
$G_3$, \Gdef, in the starting IIB theory after T-duality. ${\hat{\tilde F}}_{5(yx)}$,
arises from ${\tilde F}_3$ with no leg along the $x,y$ directions, ${\hat{\tilde F}}_{3(x)}$,
${\hat{\tilde F}}_{3(y)}$, from ${\tilde F}_3$ with a leg along either $y$ or $x$ 
direction and 
${\hat F}$ from ${\tilde F}_3$ with a leg along both $x,y$ directions. 
${\hat {\cal H}}_3$  arises from ${\cal H}_3$ with no leg along the $x,y$ directions.
Finally, the last two terms in \ibtdul, are components of  the antisymmetrised spin
 connection in the dual theory. They  arise from ${\cal H}_3$ with one leg along 
the $x$ or $y$ directions. In this context it is worth noting that the last two terms
in \ibtdul\ can be rewritten as:
\eqn\recast{{\hat j}_{xx} \eta^x \wedge  d{\hat j}_{(x)}
+ {\hat j}_{yy} \eta^y \wedge d{\hat j}_{(y)} = {\hat k}_{(x)} \wedge F^x + {\hat k}_{(y)}
\wedge F^y,}
where ${\hat k}_{(x,y)}$ are the Killing one forms defined in  \etakillings\ 
and $F^{x,y}$ are the field strengths of the associated KK gauge fields \foot{ 
This follows from the standard relation between the KK gauge potential and the metric,
$A_{\mu}^{x}={\hat j}_{\mu \alpha} {\hat j}^{\alpha x}, \alpha=x,y,$ and similarly for 
$A_{\mu}^y$.}. 

\medskip
\noindent
{\it More on the ACS in the two T-duality case}
\medskip

Before proceeding let us discuss  the ACS defined in the dual IIB theory in more detail.
In general  for the case  after two T-dualities
we find that the ACS is not integrable. Let us illustrate this lack of integrability
 with the example 
discussed in  section 3.1, which was also used in the  discussion of the ACS in the 
one T-duality case, \condint\ onwards, above. 

Starting in the IIB theory with a metric of form $ds^2=dz^id{\bar z}^{\bar i}$, we now T-dualize along both $x^1,y^1$. The three holomorphic forms are then given by 
\eqn\holtt{\eta^1_H=\eta^x+\tau^1 \eta^y,\eta^2_H=\eta^2+\tau^2\eta^5,\eta^3_H=\eta^3
+\tau^3 \eta^6.}
Now considering $d\eta_H^1$ we find that \condint\ is not met, except at special points in
 moduli  space where $\tau^1=\tau^2$. Note, this is different from the constraint in 
moduli space imposed by supersymmetry, \modtwo, except at the special point
$\tau^1=\tau^2=i$. Thus, generically, we see that the ACS is not integrable.

One finds that an alternative ACS can be defined which is integrable in this example. 
It corresponds to keeping the definition of the holomorphic one form
$\eta_H^1, \eta_H^3$ the same as in \holtt, but changing 
\eqn\chngholt{\eta^2_H=\eta^2 + \tau^1 \eta^5.} 
Now it is easy to see that the integrability condition \condint\ is met. 

We have not explored the question of whether an integrable ACS exists in full generality.
However, we expect  in several cases, obtained for example after two T-dualities,
that  this is true \TnT. 
If ${\hat \epsilon}$ is the spinor in the dual IIB theory that meets the 
susy conditions, we write 
\eqn\spiltspin{{\hat \epsilon}={\hat \epsilon}_{+} +  {\hat \epsilon}_{-}}
where ${\hat \epsilon}_{\pm}$ are spinors of definite $SO(6)$ chirality \foot{
Note after two T-dualities the spinor ${\hat \epsilon}$ is not of B-type.}. 
An alternate ACS can then be defined, using ${\hat\epsilon}_{+}$, of the form 
\eqn\altacs{J^a_b=({\hat \epsilon}_{+})^{\dagger}[\gamma^a,\gamma_b] ({\hat \epsilon}_{+})}
and similarly, using ${\hat \epsilon}_{-}$.  These alternate ACSs are 
 expected to be integrable in several instances. 

In the  example discussed above for instance,
the ACS \altacs\ is related to the one used in our susy analysis  by a rotation in the 
$\eta^2,\eta^5$ plane by $\pi/2$. This takes $\tau_2 \rightarrow -1/\tau_2$.
In view of
the constraint \modtwo, this means that the ACS \altacs\ in this example
is the same
as the ACS \chngholt, which we argued above is integrable. 
The definition of the ACS \altacs\ is also analogous to that of 
Strominger \Andy\ (see also \DasBeck).

\subsec{Superpotential}

In this section we will first construct a superpotential in the IIA
theory obtained by one T-duality in \S4.1 above, which correctly
implements the $(2,1)$ condition on the flux $G_{IIA}$. Next we will
verify that in the example of \S3.1\ the superpotential gives the
correct tensions for various BPS branes in the theory, and the correct
constraints on moduli.  We then briefly present a similar analysis for
the IIB theory obtained after two T-dualities (as described in \S4.2).
Throughout this subsection, we work in the approximation where the
warp factor \warpeq\ has been set equal to one.

\medskip
\noindent
{\it One T-Duality: The IIA Superpotential}
\medskip

Our discussion in this section  makes use of the ACS introduced in \S4.1\
\defacsa,\holoeta. The ACS is parametrized by a period matrix $\tau^i_j$.
One can define a holomorphic $(3,0)$ form (with respect to this ACS), which is given by 
\eqn\defomega{\Omega=\eta_H^1\wedge \eta_H^2\wedge \eta_H^3.}
Note that due to the torus being twisted, $\Omega$ is not closed in general.

The superpotential which implements the $(2,1)$ condition for
$G_{IIA}$ is given by
\eqn\defsupa{W_{IIA}=\int_M G_{\rm IIA} \wedge \Omega_{\rm IIA}.}
The argument leading to \defsupa\ is closely analogous to that in the
original IIB theory.  We begin by noting that $\Omega$ depends on
period matrix, $\tau^i_j$. Varying $\Omega$ with repect to $\tau^i_j$
gives us a basis of nine $(1,2)$ forms which are well defined on the
twisted torus \foot{By well-defined we mean that they have the
required periodicity properties imposed by the non-trivial twists. In
general, any form which has constant coefficients when expanded in the
$\eta$ basis \defetaspin, \defetaspinb\ meets this condition.}.  Thus
demanding that
\eqn\extrone{\partial W_{IIA}/\partial {\tau^i_j}=0,}
imposes the condition that $G_{IIA}$ has no $(1,2)$ terms in it \foot{
We assume here that to begin with $G_{IIA}$ has constant coefficients
in the $\eta$ basis. This is manifestly true for the expression
\gintwofinal\ obtained by T-duality.}.

Next, we note that $G_{IIA}$ depends on the field \foot{This is in
fact the dilaton of the original IIB theory rewriten in the IIA
variables.  It has a partner, which in IIA language is $A_{1(x)}$.
Dependence on the partner arises, e.g., from ${\tilde F_{4}}$.}
$\sqrt{g_{xx}}/g_s^{IIA}$.  Varying with respect to it and demanding
that $W_{IIA}=0$ then sets the $(3,0)$ and $(0,3)$ terms of $G_{IIA}$
to vanish, yielding the required result that $G_{IIA}$ is of type $(2,1)$.

One more comment is in order. 
Notice that $G_{IIA}$ has an explicit dependence on the metric component
$g_{xx}$ as well
(in the terms proportional to $F_2$ etc). However, varying with respect to it does 
not give any additional constraints.\foot{The $\tau^x_j$ components 
 of the period matrix always come multipled by $g_{xx}$ in $W_{IIA}$.}

\medskip
\noindent
{\it Brane Tensions in the IIA theory}
\medskip

As described above, the primitive (2,1) ISD condition can be imposed
by a superpotential $\int G\wedge\Omega$, supplemented by the
condition $J\wedge G = 0$.  So, we have
\eqn\IIAsuper{W_{\rm IIA} = \int G_{\rm IIA} \wedge \Omega_{\rm IIA},}
where $G_{\rm IIA}$ is given by Eq.~\gintwofinal.  As a check that this
is the correct expression, we can compute the tensions of domain walls
from wrapped branes, and verify that these tensions satisfy the
formula
\eqn\DeltaW{T_{\rm domain\ wall} = a \bigl| \Delta W\bigr| 
= a \biggl| \int \Delta G\wedge\Omega\biggr| ,}
with the appropriate choice of $a$.\foot{We will not determine $a$
from first principles here, but it is possible to do so by paying
proper attention to normalization.  The tension in Einstein frame is
then $e^{-{\cal K}/2}\Delta W$, where ${\cal K}$ is the K\"ahler
potential.  One then needs to convert to string frame to compare to
the results of this section.}  For simplicity, we restrict to the case
of a $T^6$ that factorizes into $T^2\times T^2\times T^2$ with respect
to both the K\"ahler and complex structure moduli:
\eqn\diagdz{dz^i = dx^i + \tau^i dy^i,\quad \tau^i =
iR_{y^i}/R_{x^i}\quad \hbox{(no sum)},}
\eqn\diagJ{J=\sum \rho_i dz^i dz^{\bar\imath},\quad
\rho_i = iR_{x^i}R_{y^i}\quad \hbox{(no sum)}.}
We also assume that the IIB dilaton-axion has no real (axionic)
component.  It is convenient to define the quantity
\eqn\OmegaIIB{\tilde\Omega_{\rm IIB} = R_{x^1}R_{x^2}R_{x^3} \Omega
= (R_{x^1}dx^1 +
iR_{y^1}dy^1)\wedge(1\rightarrow2)\wedge(1\rightarrow3).}
The utility of this definition is that the components of
$\tilde\Omega_{\rm IIB}$ are the powers of $i$ times the volume forms
of various three-cycles.  In terms of $\tilde\Omega_{\rm IIB}$,
\eqn\newDeltaOmega{T^{\rm IIB}_{\rm domain\ wall} = c_{\rm IIB} \biggl|
\int\Delta G_{\rm IIB}\wedge \tilde\Omega_{\rm IIB}\biggr|, 
\quad c_{\rm IIB} = a_{\rm IIB}/(R_{x^1}R_{x^2}R_{x^3}).}
Similarly, in the dual IIA theory obtained by T-dualizing in the
$x$-direction, it is convenient to define an analogous quantity
$\tilde\Omega_{\rm IIA}$, obtained from \OmegaIIB\ by replacing $R_x
dx$ with $R_x^{-1}\eta^x$.  We then obtain a IIA expression for the
domain wall tension, of the same form as \newDeltaOmega.  The IIA and
IIB expressions for the tension must be equal.

Consider first a D5-brane that wraps a three-cycle $\gamma_3$ of $T^6$
on IIB side.  Then, as we pass through the resulting domain wall in
${\bf R}^{3,1}$, we have $\Delta G_{\rm IIB} = \Delta F_3$, where
$F_3/(2\kappa_{10}^2\mu_{D5})$ is in the cohomology class Poincar\'e
dual to $\gamma_3$.  Consequently,
\eqn\Tdfive{c_{\rm IIB}\int\Delta G_{\rm IIB}\wedge {\tilde\Omega}_{\rm IIB}
= c_{\rm IIB} 2\kappa_{10}^2\mu_{\rm D5}\int_{\gamma_3}{\tilde\Omega}_{\rm IIB}
= c_{\rm IIB} (2\kappa_{10}^2 g_s^{\rm IIB})\tau_{\rm
D5}\int_{\gamma_3} {\tilde\Omega}_{\rm IIB}.} 
Eqs.~\newDeltaOmega\ and \Tdfive\ correctly give the tension of the
domain wall if
\eqn\IIBpropconst{1/c_{\rm IIB} = 2\kappa_{10}^2 g_s^{\rm IIB},}
and if the pullback of ${\tilde\Omega}_{\rm IIB}$ to $\gamma_3$ is a phase
factor times the volume form.  It is easy to see from the explicit
form of $\tilde\Omega$ \OmegaIIB, that if the latter is true, then
\eqn\DsixCase{\tilde\Omega_{\rm IIA}\wedge R_x^{-1}\eta^x = e^{i\alpha}
{\rm Vol}(\gamma_4)
\qquad\hbox{($\gamma_4$ wrapped by T-dual D6)\quad D5 $\perp x$},}
\eqn\DfourCase{R_x \bigl(\tilde\Omega_{\rm IIA}\bigr)_{xij}\half
dx^i\wedge dx^j = e^{i\alpha} {\rm Vol}(\gamma_2)
\qquad\hbox{($\gamma_2$ wrapped by T-dual D4)\quad D5 $\parallel
x$},}
where $e^{i\alpha}$ is a phase factor and Vol$(\gamma)$ denotes the
volume form on $\gamma$.  In these two cases we also have
\eqn\FtwoEq{g_{x[i}F_{jk]}{\textstyle{1\over3!}}dx^i\wedge dx^j\wedge
dx^k = R_x^{-2} \eta^x\wedge F_{(2)},}
\eqn\FfourEq{F_{xijk} \eta^x\wedge {\textstyle{1\over3!}}dx^i\wedge
dx^j\wedge dx^k = F_{(4)},}
respectively, so that $G_{\rm IIA}\wedge\tilde\Omega_{\rm IIA} = R_x^{-1}
F_{(2)}\wedge {\rm Vol}(\gamma_4)$ in the first case, and $G_{\rm
IIA}\wedge\tilde\Omega_{\rm IIA} = R_x^{-1} F_{(4)}\wedge {\rm
Vol}(\gamma_2)$ in the second.  Then, from
\eqn\DiracQuant{{1\over2\kappa_{10}^2\mu_{\rm D4,D6}}\int_{\gamma_{2,4}}
F_{(2),(4)} = \mu_{\rm D4,D6} = \tau_{\rm D4,D6}/g_s^{\rm IIA},}
we have
\eqn\IIAdeltaW{\eqalign{c_{\rm IIA}\int\Delta G_{\rm IIA}\wedge
\tilde\Omega_{\rm IIA} &= c_{\rm IIA} R_x^{-1}
(2\kappa_{10}^2g_s^{\rm IIA})\tau_{\rm D4,D6}\int_{\gamma_{2,4}} {\rm
Vol}(\gamma_{2,4})\cr &= c_{\rm IIA}(2\kappa_{10}^2g_s^{\rm IIA}/R_x) T^{\rm
IIA}_{\rm domain\ wall}.}}
This is a nontrivial consistency check, since we obtain the same
proportionality constant
\eqn\IIApropconst{1/c_{\rm IIA} = 2\kappa_{10}^2 g_s^{\rm IIA}/R_x,}
in either case.

Similarly, for a wrapped NS5-brane, 
\eqn\IIBNS{\eqalign{%
c_{\rm IIB} \int \Delta G_{\rm IIB}\wedge{\tilde\Omega}_{\rm IIB} 
&= c_{\rm IIB}{i\over g_s^{\rm IIB}}\int {\cal
H}\wedge{\tilde\Omega}_{\rm IIB} = c_{\rm IIB} {i\over g_s^{\rm IIB}}
\Bigl(2\kappa_{10}^2\bigl(g_s^{\rm IIB}\bigr)^2\Bigr)
\tau_{\rm NS5}\int_{\gamma_3}{\rm Vol}(\gamma_3)\cr &= c_{\rm IIB}
\bigl(2i\kappa_{10}^2 g_s^{\rm IIB}\bigr) T^{\rm IIB}_{\rm domain\
wall}.}}
If the NS5-brane does wrap the $x$-direction, then $H_{ijk}$ in
IIA is equal to ${\cal H}_{ijk}$ in IIB, and
\eqn\IIANSperp{\eqalign{c_{\rm IIA}\int\Delta G_{\rm
IIA}\wedge\tilde\Omega_{\rm IIA} &= c_{\rm IIA} {i\over g_s^{\rm IIB}}
\Bigl(2i\kappa_{10}^2 \bigl(g_s^{\rm IIA}\bigr)^2\Bigr) T^{\rm
IIA}_{\rm domain\ wall}\cr &= c_{\rm IIA} \bigl(2i\kappa_{10}^2
g_s^{\rm IIA}/R_x\bigr) T^{\rm IIA}_{\rm domain\
wall},}}
where in the last step we have used the T-duality rule $g_s^{\rm IIA} =
g_s^{\rm IIB}/R_x$.

If the NS5-brane does not wrap the $x$-direction, then it is possible to
show by a calculation analogous to \dualH\ that
\eqn\IIANSpar{\bigl(\omega_{(3)}\bigr)_{ijx} = R_x^{-2} {\cal H}_{ijx},
\qquad \bigl(\omega_{(3)}\bigr)_{ijk} = {\cal H}_{ijk} = 0\ {\rm
otherwise}.}
Eqs.~\IIANSperp\ and \IIANSpar\ together give $c_{\rm IIA}= c_{\rm
IIB} R_x^2$, which agrees with Eqs.~\IIApropconst\ and \IIBpropconst.
The carrier of $\bigl(\omega_{(3)}\bigr)_{ijx}$ charge is a
Kaluza-Klein monopole, which can be regarded as a KK5-brane, or
twisted 6-brane \hull.

\medskip
\noindent
{\it Moduli constraints from the IIA superpotential}
\medskip
In the example considered in \S3.1\ it is
 possible to directly derive the constraints on moduli \modonet\ and
\modtwot\ which apply in the T-dual IIA description, by computing
and stationarizing the superpotential \IIAsuper.  
Direct calculation of $G_{\rm IIA}$ \gintwofinal\ using 
the formulae in App.~A
yields
\eqn\gis{\eqalign{{1\over 2} G_{\rm IIA} &= 
dy^1 \wedge dy^2 \wedge dy^3 + {\tilde R_{x^1}^2}
(dx^1 + x^2 dx^3) \wedge dx^2 \wedge dy^3\cr 
& \phantom{=}-\bigl(i/g_s^{\rm IIA}\bigr) \tilde R_{x^1}
\left( dy^1 \wedge dy^2 \wedge dx^3 + 
\tilde R_{x^1}^2 dx^1 \wedge dx^2 \wedge dx^3 \right)~.}}
So one finds that $W_{\rm IIA}$ is
\eqn\wiiais{ {1\over 2} W_{\rm IIA} ~=~1 + 
\rho^1 \tau^2 + \bigl(\tilde R_{x^1}/g_s^{\rm IIA}\bigr)
(\tau^3 + \rho^1 \tau^2 \tau^3)~.} 
Here the $\rho$ and $\tau$ moduli are being defined as in a
conventional square six-torus, so e.g.  $\rho^1 = i \tilde R_{x^1}
R_{y^1}$ and so forth.\foot{As a reminder, $\tilde R_{x^1}$ is the
radius of the $x^1$ circle in the IIA theory.  The other radii are the
same as in the original IIB theory and are denoted without tildes.}

As an aside, we should note that it is not surprising that $1/g_s^{\rm
IIA}$ appears in the superpotential together with a factor of the IIA
radius $\tilde R_{x^1}$.  The 10d IIB axio-dilaton $\phi$ is a complex
scalar, and can appear in a 4d ${\cal N}=1$ chiral multiplet, and
hence in a superpotential.  In contrast, the 10d IIA dilaton is a {\it
real} scalar, and first becomes part of a complex scalar in 9d; and
the combination that appears in $W_{\rm IIA}$ is precisely (by
T-duality) the IIB dilaton $\phi$.  This is the field that appears as
part of a chiral multiplet in the low-energy 4d supersymmetric theory.

At this point, we see immediately that the constraints on moduli
that we will
find from varying \wiiais\ will agree with the results \modonet\ and \modtwot\
that follow by dualizing the original IIB constraints.  This is because 
\wiiais\ precisely agrees (up to the change of notation from IIB to IIA
variables) with
the original IIB superpotential.

\medskip
\noindent{\it Superpotential after Two T-Dualities}
\medskip

After two T-dualities, the superpotential is
\eqn\twoTsup{\hat W_{\rm IIB} = \int\hat G_3\wedge \hat\Omega_{\rm
IIB},}
where ${\hat G}_3$ is given by Eq. \ibtdul. An argument similar to that of \S4.3\
shows that it correctly implements the $(2,1)$ condition on the flux, ${\hat G}_3$.
A similar identification of
domain wall tensions can be performed again in this case.  The terms
in $\hat G_3$ and the corresponding domain walls are summarized in the
following table:

\bigskip\vbox{\baselineskip=14pt
\settabs\+\indent\indent&Term in $\hat G_3$\qquad\qquad\qquad&Domain Wall\cr
\+&Term in $\hat G_3$&Domain Wall\cr
\smallskip
\+&$F_{5(yx)}$&D3-brane transverse to $x$ and $y$&\cr
\+&$k_{(x)}\wedge F_{3(y)}$&D5-brane wrapped on $x$, transverse to $y$&\cr
\+&$k_{(y)}\wedge F_{3(x)}$&D5-brane wrapped on $y$, transverse to $x$&\cr
\+&$\bigl(\detjht\bigr)\eta^x\wedge\eta^y\wedge F_1$&D7-brane wrapped
on $x$ and $y$&\cr
\+&$\hat{\cal H}_3$&NS5-brane wrapped on $x$ and $y$&\cr 
\+&${\hat j}_{xx}\eta^x\wedge d{\hat j}_{(x)}$&KK5-brane transverse to $x$\cr
\+&${\hat j}_{yy}\eta^y\wedge d{\hat j}_{(y)}$&KK5-brane transverse to $y$\cr
}

\newsec{Discussion}

The main lesson we can draw from this paper is that new solutions of
the string equations of motion based on non Calabi-Yau geometries exist.
These can provide supersymmetric vacua after suitable flux is turned on. 
While we have concentrated on the type II theories in this paper, it is
clear (for example by duality, as in \DasBeck) 
that similar constructions exist in the heterotic theory. 
Moreover, in the simplest examples described in this paper, we 
can construct a superpotential sensitive
to this departure from ``Calabi-Yauness.''  It would be very interesting
to characterize properly in mathematical terms what the class of
such compactifications is, and discover how to most efficiently parametrize
their moduli spaces and superpotentials.  Other recent work in this
direction has appeared in \refs{\Janetal,\Humbgrp}.   

There are also several connections of our new vacua with
other backgrounds.  As we saw in \S3, in some cases, we expect our models
to have a lift to an M-theory compactification on a manifold of special
holonomy.  It would be very interesting to find examples where the
M-theory Calabi-Yau (for ${\cal N}=2$) or $G_2$ (for ${\cal N}=1$) geometry
could be specified more explicitly. 
Perhaps one can go further towards determining the Calabi-Yau $X$
which appeared in \S3.1\ by gluing together Atiyah-Hitchin and ALE metrics.

Another connection arises by using the duality explored by Hull in
\massive.  There, it is argued that M-theory on the nilmanifold is
dual to the massive type IIA theory.  It follows from this
that type IIA theory compactified on the nilmanifold should be
dual to appropriate compactifications of the massive IIA theory.
But we have seen in \S3.1\ that compactifications of the IIA theory
on the nilmanifold (and an additional transverse $T^3$), with appropriate
orientifold actions and background fluxes, yield supersymmetric vacua.   
This implies that the massive IIA theory, which has no
ground state with unbroken supersymmetry in ten dimensions, 
does have vacua with unbroken supersymmetry in lower dimensions.

There are many clear directions for future work along the lines of
this paper:

\noindent
$\bullet$ We have written down examples of novel geometries which arise
by dualizing known vacua.  However, the superpotential formulae of \S4
can be applied directly to find supersymmetric vacua.  It would be
interesting to exhibit a vacuum which is not related by duality 
to a compactification
on a Calabi-Yau orientifold or $G_2$ space in {\it any~} controlled limit.

\noindent
$\bullet$ Our construction started from the $T^6/Z_2$ orientifold, and
the simplicity of $T^6$ is what is responsible for the simple nature of
our dual metrics and superpotential formulae.  However, it is clear that
starting from a more general Calabi-Yau, one could still
try to T-dualize along the $T^3$ fibers \SYZ.  As long as $H$ does not
have any components with more than one leg along the fibers, this should
yield a sensible dual geometry.  It would be worthwhile to explore
the geometries that arise in this way.  Some steps in this direction
have been taken in the recent papers \refs{\Janetal,\Humbgrp}. 

\noindent
$\bullet$ The detailed structure of the 
moduli spaces of these new compactifications are somewhat
mysterious.  In the large volume limit, the moduli space should 
have a description in geometrical terms (analogous to the description
of Calabi-Yau moduli space which follows from Yau's theorem).
It would be nice to work out an intrinsic formulation of this moduli
problem in terms of large radius geometry. 
In the examples of this paper, and possibly more generally, it could be that
the correct procedure is to view all of these vacua as being specified
by: i) a choice of Calabi-Yau manifold $M$, ii) a choice of background
NS and RR fluxes on $M$, iii) a choice of background ``KK monopole flux''
or ``twisting'' on $M$ (in the examples at hand, this corresponds to 
the choice of twisting in the nilmanifold).  One then stationarizes
the superpotential (which now contains a contribution from the twisting
iii)), and obtains in this way constraints on the ``formerly Calabi-Yau''
moduli.  However this prescription, in addition to being vague, 
depends on various conjectures
(about the existence of metrics with appropriate windings iii) given the
Calabi-Yau metric on $M$, for instance) which are not obviously true.

\noindent
$\bullet$ We have seen in this paper that compactifications with flux,
after T-duality, turn into twisted tori. The twisting gives rise to a 
non-compact group and the twisted torus can be viewed as a coset of this 
non-compact 
group. In the examples studied here these turn out to be 
generalizations of the nilmanifold. 
This raises the natural question of 
whether all Scherk-Schwarz compactifications
can be understood in 
this way as cosets.  More generally, the connection 
between flux, twisted tori and 
cosets (or orbifolds) needs 
to be understood better, building on the recent work in \Atish. 

\noindent
$\bullet$ Finally, it would be nice to see if
the new compactifications and computable potentials described 
here can, together
with other ingredients (notable possibilities are D-brane instanton
effects and gaugino condensates), lead to models of 
interest in string
cosmology. 
A discussion of the cosmology arising from IIB Calabi-Yau
orientifolds with three-form 
flux recently appeared in \freycos, the conclusion
being that the potentials obtained to date are too steep to serve
a useful role in models of inflation or quintessence.

\appendix{A}{T-duality maps of fields}

In this appendix we express the T-duality maps from IIA to IIB and vice 
versa. 
We begin by first explaining our conventions and notation. 
We will mainly use the conventions of Polchinski \joeP\ 
for the supergravity fields (with the same normalisations).
In particular,
\eqn\aaone{{\tilde F}_4 = dC_3 + A_1 \wedge H_3 ~,~
   {\tilde F}_5 = dC_4 - {1\over 2} C_2 \wedge {\cal H}_3
                      + {1\over 2} {\cal B}_2 \wedge F_3,}
stand for the gauge invariant RR 4-form and 5-form field strengths in the IIA and IIB theories.
Also we will use the notation
\eqn\aagf{{\tilde F}_3 = F_3 - C_0 {\cal H}_3 ~,~}
in the IIB theory. 

We also remind the reader of some of our notation.  $x$ denotes the direction along which we T-dualise.
 $\mu,\nu$ denote corrdinate indices and take six values, 
$\alpha,\beta$ denote  all coordinate indices except $x$ and take five values.
Finally, $a,b,$ denote tangent space indices. 
 
To save clutter we denote the metric as $j_{\mu\nu}$ and $g_{\mu\nu}$
in the  IIB and IIA theories respectively. 
We also denote the NS two form potential and field strengths as 
${\cal B}, {\cal H}_{3}$ and $B, H_{3}$ in the IIB and IIA theories respectively.

Given an $n$ form $F_{n}$ we use the notation $F_{n(x)}$ to denote
an $n-1$ form whose components are given by:
\eqn\defnx{[F_{n(x)}]_{i_1, \cdots i_{n-1}}=[F_n]_{xi_1 \cdots i_{n-1}}.}
Finally  
\eqn\defjxgx{j_{(x)} = {1\over j_{xx}} j_{x\alpha} dx^{\alpha}~,~
 g_{(x)} = {1\over g_{xx}} g_{x\alpha} dx^{\alpha} }
denote one forms in IIB and IIA theory and 
\eqn\defspin{\omega_{(x)} = - dg_{(x)}~, }
denotes a two form in IIA theory.

With these definitions, the map between IIA, and IIB theory under 
T-duality in the $x$-direction is given as follows \refs{\BHO,\HassanBV}:
The Neveu-Schwarz fields transform as 
\eqn\twoab{\eqalign{
& g_{xx} = {1\over j_{xx}} \cr
& g_{x\alpha} = - {{\cal B}_{x\alpha}\over j_{xx}} \cr   
& g_{\alpha\beta} = j_{\alpha\beta} 
                 - {1\over j_{xx}} (j_{x\alpha}j_{x\beta}
- {\cal B}_{x\alpha}{\cal B}_{x\beta})\cr
& B_{x\alpha} = - {j_{x\alpha}\over j_{xx}} \cr
& B_{\alpha\beta} = {\cal B}_{\alpha\beta} 
                - {1\over j_{xx}} (j_{x\alpha} {\cal B}_{x\beta}
                - {\cal B}_{x\alpha} j_{x\beta}) \cr
& g_s^{IIA} = {g_s^{IIB}\over{\sqrt{j_{xx}}}}
}}
Here the left hand refers to fields in the IIA theory and the right hand side 
to fields in the IIB theory. 
For the three form field strength, $H_3$ this takes the form,
\eqn\fstrength{\eqalign{
& H_{3(x)} =  dj_{(x)} \cr
& H_3 = {\cal H}_3 - {\cal H}_{3(x)} \wedge j_{(x)} 
                     - {\cal B}_{(x)} \wedge dj_{(x)} 
}}
The Ramond fields transform as
\eqn\rfields{\eqalign{
& F_{2(x)} = F_1  \cr
& F_2 =  {\tilde F}_{3(x)}  - {\cal B}_{(x)} \wedge F_1 \cr
&{\tilde F}_{4(x)} = {\tilde F}_3  - j_{(x)} \wedge {\tilde F}_{3(x)} \cr
&{\tilde F}_4 = {\tilde F}_{5(x)} 
- {\cal B}_{(x)} \wedge \left( {\tilde F}_3  
- j_{(x)} \wedge {\tilde F}_{3(x)} \right)
}}

In the formulae above, a field strength with and without a leg along the $x$ direction are 
 denoted as $F_{n(x)}$, and $F_n$ respectively. 

The inverse of these expressions is given by
\eqn\toremove{\eqalign{
& j_{xx} = {1\over g_{xx}} \cr
& j_{x\alpha}  = - {B_{x\alpha}\over g_{xx}} \cr
& j_{\alpha\beta} = g_{\alpha\beta} 
                   - {1\over g_{xx}} (g_{x\alpha}g_{x\beta}
                   - B_{x\alpha}B_{x\beta})\cr
& {\cal B}_{x\alpha} = - {g_{x\alpha}\over g_{xx}} \cr
& {\cal B}_{\alpha\beta} = { B}_{\alpha\beta} 
                - {1\over g_{xx}} (g_{x\alpha} {B}_{x\beta}
                - {B}_{x\alpha} g_{x\beta}) \cr
& g_s^{IIB} = {g_s^{IIA}\over{\sqrt{g_{xx}}}}
}}
\eqn\twoba{\eqalign{
& {\cal H}_{3(x)}  = - \omega_{(x)} \cr
& {\cal H}_3 = H_3  + \omega_{(x)} \wedge B_{(x)} 
                    - g_{(x)} \wedge H_{3(x)} \cr
}}
and 
\eqn\invforf{\eqalign{
& F_1 = F_{2(x)} \cr
& {\tilde F}_{3(x)}  = F_2 - g_x \wedge F_{2(x)} \cr
& {\tilde F}_3 = {\tilde F}_{4(x)} 
                 - B_{(x)} \wedge \left( F_2 - g_x \wedge F_{2(x)} \right) \cr
& {\tilde F}_{5(x)} = {\tilde F}_{4} - g_{(x)} \wedge {\tilde F}_{4(x)} \cr
}}

\appendix{B}{T-duality map of spinor conditions} 
\def\i{\alpha}
\def\j{\beta}
\def\k{\gamma}
In this appendix we discuss the  T-duality transformation rules for  the 
IIB and IIA spinor conditions.
Our discussion largely follows \HassanBV.
We first discuss the map for  T-duality along one direction and then, more briefly,
two directions.

\medskip
\noindent{\it One T-duality} 
\medskip

We denote the spinors and vielbein  in the IIB (IIA) theory with (without) a tilde superscript.
The vielbein in IIB and IIA theory are related by
\eqn\testq{\tilde{e}^{\mu}_a = Q^{\mu}_{\nu} e^{\nu}_a }
where the matrix $Q$ is given by
\eqn\matr{
Q = \left(
\matrix{ & g_{xx} &  (g + B )_{x\i} \cr
        & 0      &  {\bf 1} }
\right)
}

Gamma matrices with tangent space indices  are the same in the IIB and IIA theories, i.e.
\eqn\relgamma{{\tilde \gamma}^{a} = \gamma^a~.}
We will denote these as $\gamma^a$ below. 

One can show \HassanBV\ that if
\eqn\abone{{\tilde \epsilon}={\tilde \epsilon_L} + i {\tilde \epsilon_R},}
meets the spinor conditions imposed by susy in the IIB theory, then  
\eqn\abtwo{\epsilon= \epsilon_L + i \epsilon_R}
meets the spinor conditions in the IIA theory.
Here
\eqn\defelab{{\tilde \epsilon_L}= \Omega \epsilon_L,}
\eqn\deferab{{\tilde \epsilon_R}= \epsilon_R,}
and
\eqn\defomegab{\Omega = - {1 \over \sqrt{g_{xx} }} \gamma_{11} \gamma_9 ~.}
The above statement is true if the IIB and IIA fields are related under 
T-duality by the transformations \twoab, \twoba.

We now turn to mapping the spinor conditions of type IIB to type IIA.  
We will not demonstrate the map in full detail.
Instead we give one representative example.  
Consider the dilatino variation in the IIB theory. 
Since the dilaton-axion is constant in the IIB theory, it is
(eq.(2.3) of \GPOne ):
\eqn\apbdila{ \delta\tilde\lambda = - {i\over 24} {\tilde G} \tilde\epsilon,} 
where 
\eqn\apbsone{ {\tilde G} = \tilde\gamma^{\mu \nu \sigma }
{\left[G_{3}\right]}_{\mu \nu \sigma}} 
 with 
\eqn\apbstwo{\eqalign{
\left[G_{3}\right]_{\mu \nu \sigma } 
\equiv & \left[F_3\right]_{\mu \nu \sigma } 
- \phi \left[{\cal H}_3\right]_{\mu \nu \sigma }  \cr
= & \left[{\tilde F}_3\right]_{\mu \nu \sigma } 
- \left(i/ g_s^{IIB}\right) \left[{\cal H}_3\right]_{\mu \nu \sigma}}}
and 
$\tilde\epsilon = \tilde\epsilon_L + i \tilde\epsilon_R ~. $
Then using the relation between the vierbeins \testq, we have the relations:
$$\tilde\gamma^x = g_{xx} \gamma^x 
+ \left[g + B\right]_{x\alpha} \gamma^{\alpha} $$
and
$$\tilde\gamma^{\alpha} = \gamma^{\alpha}~. $$

Using the T-duality relations between the fields \twoba, the flux
\eqn\pbfluxha{\left[{\cal H}_3\right]_{\i\j\k} 
= \left[H_3\right]_{\i\j\k} 
+ \left[\omega_{(x)} \wedge B_{(x)}\right]_{\i\j\k} 
- \left[g_{(x)} \wedge H_{3(x)}\right]_{\i\j\k}} 
\eqn\pbfluxhb{\left[{\cal H}_3\right]_{x\i\j} = - \omega_{x\i\j}. }
So 
\eqn\abfinalh{ \tilde\gamma^{\mu\nu\rho}\left[{\cal H}_3\right]_{\mu\nu\rho}
= \left[ H_3 - g_{xx} g_{(x)} \wedge \omega_{(x)} 
   - g_{(x)} \wedge H_{3(x)}\right]_{\i\j\k} \gamma^{\i\j\k}
   - g_{xx} \gamma^{x\i\j}\omega_{x\i\j}. }
As mentioned at the end of this appendix, the RHS can be written as follows:
\eqn\abfinalheta{\tilde\gamma^{\mu\nu\rho}\left[{\cal H}_3\right]_{\mu\nu\rho}
= \left[ H_3 - g_{xx} \eta^x \wedge \omega_{(x)}
  - \eta^x\wedge H_{3(x)} \right]_{\mu\nu\rho} \gamma^{\mu\nu\rho}.}

Similarly, using \twoba,  
$$ \left[{\tilde F}_3\right]_{\i\j\k} 
=  \left[{\tilde F}_4\right]_{x\i\j\k} 
- \left[B_{(x)} \wedge F_2\right]_{\i\j\k}
+ \left[ B_{(x)} \wedge g_{(x)} \wedge F_{2(x)}\right]_{\i\j\k} $$
and
$$ \left[{\tilde F}_3\right]_{x\i\j} 
= \left[F_2 - g_{(x)} \wedge F_{2(x)}\right]_{\i\j}~. $$
So we find
\eqn\forfth{\eqalign{
 \tilde\gamma^{\mu\nu\rho}\left[{\tilde F}_3\right]_{\mu\nu\rho}
= & \left[ {\tilde F}_{4(x)} + g_{xx} g_{(x)} \wedge F_2\right]_{\i\j\k} 
\gamma^{\i\j\k} + g_{xx} \left[F_2 
- g_{(x)} \wedge F_{2(x)}\right]_{\i\j} \gamma^{x\i\j}. 
}}
Once again as discussed at the end of the appendix this can be expressed as
\eqn\abfinalfeta{\tilde\gamma^{\mu\nu\rho}
\left[{\tilde F}_3\right]_{\mu\nu\rho}
= \left[{\tilde F}_{4(x)} 
+ g_{xx} \eta^x \wedge F_2\right]_{\mu\nu\rho}\gamma^{\mu\nu\rho}.}

From \abfinalheta, \abfinalfeta\ we obtain
\eqn\fortdulg{
\tilde\gamma^{\mu \nu \sigma }\left[G_{3}\right]_{\mu \nu \sigma }
= \gamma^{\mu \nu \sigma }\left[G_{IIA}\right]_{\mu \nu \sigma },}
where $G_{\rm IIA}$ is a three-form in the IIA theory given by
\eqn\gintwoa{G_{\rm IIA}=\left({\tilde F}_{4(x)} 
             + g_{xx} \eta^x \wedge F_2\right) 
- \left(i {\sqrt g_{xx}}\over g_s^{IIA}\right) 
\left(H_3 -g_{xx} \eta^x \wedge w_{(x)}
-\eta^x \wedge H_{3(x)}\right).}  
Setting the dilatino variation to zero in IIB sets the RHS of
\fortdulg\ to zero, yielding the condition that
\eqn\finaldilacona{\gamma^{\mu \nu \sigma }
\left[G_{IIA}\right]_{\mu \nu \sigma }{\tilde \epsilon}=0.}
The  ACS in IIA is defined by the relation \defacsa. In turn this defines a holomorphic basis,
\holoeta, in which 
\eqn\hologamma{\gamma^\ibar {\tilde \epsilon}=0.}
It is then easy to see that 
\finaldilacona\ leads to the condition that the term in $G_{IIA}$ of type $(3,0)$ vanish
and that the $(2,1)$ term satsifies the condition of primitivity \prima.

It is straightforward to repeat this discussion for the gravitino variation in IIB theory.
Mapping the resulting  conditions to IIA yield the additional constraints that 
$G_{IIA}$ is purely of type $(2,1)$ and primitive.

We end this discussion of the one T-duality case with two observations. 
First, we note that \abfinalheta\ follows from 
\abfinalh, and \abfinalfeta\ follows from \forfth, due to the fact that 
\eqn\abrel{
\left[\omega_x \wedge dx\right]_{\i\j\k} =
\left[ H_{3(x)} \wedge dx\right]_{\i\j\k} =
\left[g_{(x)} \wedge H_{3(x)}\right]_{x\i\j}
= \left[g_{(x)} \wedge \omega_{(x)}\right]_{x\i\j} = 0. }
Second, we note that the last term in the expression \gintwoa\ for
$G_{\rm IIA}$ is absent.  It arises from the term $dj_{(x)}$ in the
original IIB theory on $T^6/Z_2$, which vanishes since $j_{(x)}$ is a
constant in that theory. Thus we have \gintwofinal.

\medskip
\noindent{\it Two T-dualities}
\medskip

A similar analysis can be carried out for T-duality along two directions. 
Here we state the main conclusions in obvious notation. 

One finds that \foot{Terms proportional to ${\hat H}_{3(x)}, {\hat H}_{3(y)}$ 
are absent due to  the orientifold projection.}
\eqn\forhtt{
T_y\circ T_x~: ~~~ \left[{\cal H}_3\right]_{\mu\nu\rho} 
{\tilde\gamma}^{\mu\nu\rho}
= \left[{\hat H}_3\right]_{\ah\bh\gh} {\gamma}^{\ah\bh\gh}
+ \left[ {\hat j}_{xx} \eta^x \wedge d{\hat j}_{(x)}
+ {\hat j}_{yy} \eta^y \wedge d{\hat j}_{(y)} \right]_{\mu\nu\rho}
 {\hat\gamma}^{\mu\nu\rho}
}
and for the Ramond-Ramond field strength
\eqn\forftt{\eqalign{
T_y\circ T_x~: ~~~ \left[{\tilde F}_3\right]_{\mu\nu\rho} 
{\tilde\gamma}^{\mu\nu\rho}
= & \left[ {\tilde F}_{5(yx)} + \left( {\hat j}_{xx} \eta^x
          + {\hat j}_{xy} \eta^y\right) \wedge {\hat{\tilde F}}_{3(y)}
          - \left( {\hat j}_{xy} \eta^x + {\hat j}_{yy} \eta^y\right) 
            \wedge {\hat{\tilde F}}_{3(x)} \right. \cr
          & \left. 
          + \detjht ~ \eta^x \wedge \eta^y \wedge F_1 \right]_{\mu\nu\rho}
           {\gamma}^{\mu\nu\rho}~. 
}}
Here the indices $\ah,\bh,\hat\gamma$ run over the compact directions
except the T dualized ones $x$ and $y$ .  The above expressions imply 
that
\eqn\tyotx{
T_y\circ T_x~: ~~~ \left[G_3\right]_{\mu\nu\rho} {\tilde\gamma}^{\mu\nu\rho}
= \left[{\hat G}_3\right]_{\mu\nu\rho} {\hat\gamma}^{\mu\nu\rho}
}
where
\eqn\iibtdual{\eqalign{
{\hat G}_3 = & \left\{ {\hat{\tilde F}}_{5(yx)} + \left( {\hat j}_{xx} \eta^x 
          + {\hat j}_{xy} \eta^y\right) \wedge {\hat{\tilde F}}_{3(y)}
          - \left( {\hat j}_{xy} \eta^x 
          + {\hat j}_{yy} \eta^y\right) \wedge {\hat{\tilde F}}_{3(x)} 
          \right. \cr
          & \left. + \detjht ~ \eta^x \wedge \eta^y \wedge {\hat F}_1 \right\}
- \left(i \over g_s^{IIB}\right) \sqrt{\detjht} \left\{ {\hat{\cal H}}_3 
+ {\hat j}_{xx} \eta^x \wedge d{\hat j}_{(x)}
+ {\hat j}_{yy} \eta^y \wedge d{\hat j}_{(y)}
\right\}~. 
}}
Here we have used the definitions 
\eqn\fyxfh{\eqalign{
& \left[{\hat{\tilde F}}_{5(yx)}\right]_{\ah\bh\gh} 
= \left[{\hat{\tilde F}}_5\right]_{yx\ah\bh\gh}  \cr
& \left[{\hat{\tilde F}}_{3(x)}\right]_{\ah\bh}
= \left[{\hat{\tilde F}}_3\right]_{x\ah\bh} ~,~
 \left[{\hat{\tilde F}}_{3(y)}\right]_{\ah\bh}
= \left[{\hat{\tilde F}}_3\right]_{y\ah\bh} \cr
& \left[{\hat{ H}}_{3(x)}\right]_{\ah\bh}
= \left[{\hat{H}}_3\right]_{x\ah\bh} ~,~
 \left[{\hat{ H}}_{(y)}\right]_{\ah\bh}
= \left[{\hat{ H}}_3\right]_{y\ah\bh}
}}
and
$$\detjht = ({\hat j}_{xx} {\hat j}_{yy} - {\hat j}_{xy}^2) ~,~
{\hat j}_{(x)} = {\hat j}_{x\ah}dx^{\ah}/{\hat j}_{xx}~,~ 
{\hat j}_{(y)} = {\hat j}_{y\ah}dx^{\ah}/{\hat j}_{yy} ~. $$  

Now, in analogy with the one T-duality case, the choice of ACS discussed in \S4.2\ 
before Eq.~\fundtwo\ defines a basis of holomorphic one forms.  In this
basis one can then conclude that ${\hat G}$ is of type $(2,1)$ and
primitive.

\medskip
\centerline{\bf{Acknowledgements}}
\medskip

We would like to acknowledge very useful conversations with
A. Dabholkar, S. Gukov, S. Hellerman and G. Moore.
We are also grateful to B.~Acharya,
A.~Brandhuber, K.~Dasgupta, S.~Giddings, J.~Gomis, X.~Liu, 
L.~McAllister, S.~Mukhi, H.~Ooguri,
S.~Prokushkin, T.~Ramdas, J.~Schwarz, E.~Silverstein, H.~Verlinde and
K.~P.~Yogendran for helpful comments.  
S.K. thanks the Aspen Center for Physics for hospitality while this
work was in progress.
The work of S.K. was supported by a David and
Lucile Packard Foundation Fellowship for Science and Engineering, an
A.P. Sloan Foundation Fellowship, National Science Foundation grant
PHY-0097915, and the DOE under contract DE-AC03-76SF00515.  The work
of M.S. was supported by the DOE under contract DE-FG03-92-ER40701.
S.P.T. acknowledges support from the Swarnajayanti Fellowship, 
Department of Science and Technology, Government of India. The work of P.K.T. 
and S.P.T. was supported by the DAE. Most of all,  
P.K.T. and S.P.T. thank the people of India. 
\listrefs
\end